\documentclass[11pt]{article}
\pdfoutput=1
\usepackage{jcapmod}
\usepackage{mathtools}
\usepackage{microtype}
\usepackage[utf8]{inputenc}
\usepackage[english]{babel}
\usepackage[dvipsnames]{xcolor}
\usepackage{amsmath,amssymb,amsbsy,amstext,amsthm}
\usepackage[colorlinks=true]{hyperref}
\usepackage{microtype}
\usepackage{graphicx}
\usepackage{amsfonts}
\usepackage{upgreek}
\usepackage{exscale,relsize}
\usepackage[makeroom]{cancel}
\usepackage{soul}
\usepackage{float}
\RequirePackage{color}
\usepackage{ifpdf}
\ifpdf
\fi

\usepackage{colortbl}
\definecolor{green2}{cmyk}{0, 1, 0.5, 0}
\definecolor{lightgreen}{cmyk}{0.2, 0, 0.2, 0.2}
\definecolor{dred}{rgb}{0.9,0.2,0.5}
\definecolor{dred2}{cmyk}{0.1,0.7,0.1,0.3}
\definecolor{lightgray2}{cmyk}{0.4,0.4,0,0.8}
\definecolor{black}{cmyk}{1.0,1.0,1.0,1.0}

\AtBeginDocument{
 \hypersetup{
 urlcolor=NavyBlue,
 citecolor=NavyBlue,
 linkcolor=BrickRed,
 }
}

\allowdisplaybreaks[1]

\usepackage{colortbl}

\makeatletter
\newlength{\apb@width}
\newcommand{\autoparbox}[2][c]{\settowidth{\apb@width}{#2}\parbox[#1]{\apb@width}{#2}}

\makeatother

\numberwithin{equation}{section}

\def\beq{\begin{equation}}
\def\eeq{\end{equation}}

\def\bea{\begin{eqnarray}}
\def\eea{\end{eqnarray}}
\def\eg{{\it e.g.~}}

\def\ie{{\it i.e.~}}
\def\d{{\rm d}}

\def\d{{\rm d}}

\def\nn{\nonumber}

\def\sgm{\sigma}

\def\del{\partial}
\def\Mp{M_{\rm pl}}

\def\fr{\frac}

\def\0{{\boldsymbol 0}}

\def\fr{\frac}

\usepackage{setspace} 

\begin{document}

\begin{titlepage}

\setcounter{page}{1} \baselineskip=15.5pt \thispagestyle{empty}

\bigskip\

\vspace{1cm}
\begin{center}

{\fontsize{20}{28}\selectfont  \sffamily \bfseries {Parity Violating Non-Gaussianity from Axion-Gauge Field Dynamics
}}

\end{center}

\vspace{0.2cm}

\begin{center}
{\fontsize{13}{30}\selectfont Ogan \"Ozsoy
$^{\clubsuit}$
}
\end{center}

\begin{center}

\vskip 8pt
\textsl{
$\clubsuit$ CEICO, Institute of Physics of the Czech Academy of Sciences, Na Slovance 1999/2, 182 21, Prague.}
\vskip 7pt

\end{center}

\vspace{1.2cm}
\hrule \vspace{0.3cm}
\noindent {\sffamily \bfseries Abstract} \\[0.1cm]
We study scalar-tensor-tensor and tensor-scalar-scalar three point cross correlations generated by the dynamics of a transiently rolling spectator axion-$\mathrm U(1)$ gauge field model during inflation. In this framework, tensor and scalar fluctuations are sourced by gauge fields at the non-linear level due to gravitational interactions, providing a chiral background of gravitational waves while keeping the level of scalar fluctuations at the observationally viable levels at CMB scales. We show that the gravitational couplings between the observable sector and gauge fields can also mediate strong correlations between scalar and tensor fluctuations, generating an amplitude for the mixed type three-point functions that is parametrically larger -- $f_{\rm NL} \simeq \mathcal{O}(1-10) (r/0.01)^{3/2}$-- compared to the single-field realizations of inflation. As the amplification of the gauge field sources are localized around the time of horizon exit, the resulting mixed bispectra are peaked {\it close} to the equilateral configurations. The shape dependence, along with the scale dependence and the parity violating nature of the mixed bispectra can serve as a distinguishing feature of the underlying axion-gauge field dynamics and suggest a careful investigation of their signatures on the CMB observables including cross correlations between temperature T and E,B polarization modes.
\vskip 10pt
\hrule

\vspace{0.6cm}
 \end{titlepage}

\tableofcontents
\newpage 

\section{Introduction}

The production of primordial gravitational waves (GWs) (See \eg reviews \cite{Guzzetti:2016mkm,Caprini:2018mtu}) is a robust prediction of the inflationary paradigm \cite{Guth:1980zm,Linde:1981mu,Baumann:2009ds}.  A positive detection of such fossil GWs would therefore provide strong evidence for inflation in the early universe. The amplitude of this signal is conventionally parametrized by the so called tensor-to-scalar ratio $r$ which is a quantity targeted by a number of probes aiming to observe B-mode polarization patterns \cite{Zaldarriaga:1996xe,Kamionkowski:1996ks} in the Cosmic Microwave Background (CMB) sky \cite{Kamionkowski:2015yta}. Current limits on the tensor-to-scalar ratio from Planck and BICEP/Keck restrict $r \lesssim 0.06$  \cite{Array:2015xqh,Akrami:2018odb} and are expected to be improved by an order of magnitude by the forthcoming experiments such as CMB-S4 \cite{Abazajian:2016yjj} and LiteBIRD \cite{Hazumi:2019lys}. 

In simplest realizations of inflation based on a scalar field minimally coupled to Einstein gravity, primordial GWs originate from the quantum vacuum fluctuations of the metric amplified by the quasi-dS expansion. In this framework, the amplitude of the produced GWs is directly related to the expansion rate $H_{\rm inf}$ of the quasi-dS background, and thus if GWs are observed they would provide us the energy scale of inflation, $H_{\rm inf} \sim 10^{-5} (r/0.01)^{1/2} \Mp$ and give us the first hints on the quantum nature of gravity. 
More importantly, in this setup, the resulting GW signal is expected to posses the following properties: i) near scale invariance (with a slight red-tilt) ii) near Gaussianity and iii) parity conservation\footnote{Within the generalized scalar-tensor theories of single field inflation, a blue tilted tensor spectrum can be generated for backgrounds that exhibit a transient non-attractor era \cite{Mylova:2018yap,Ozsoy:2019slf}. On the other hand, parity violation in the tensor sector can be induced by non-minimal couplings between inflaton and the metric  \cite{Alexander:2004wk,Satoh:2008ck}.}. In order to have a firm understanding of the fundamental nature of inflation, it is therefore crucial to test the robustness of these predictions by exploring viable alternative mechanisms that can generate GWs during inflation. 

In fact, the properties i)-iii) of tensor fluctuations does not generically hold and can be invalidated if additional energetic enough field configurations present during inflation (See \eg \cite{Cook:2011hg,Senatore:2011sp}). From a top-down model building perspective, a rich particle content during inflation is not just an interesting possibility but appears to be a common outcome of many theories beyond the Standard Model of Particle Physics (See \eg \cite{Baumann:2014nda}). For example, low energy effective descriptions of string theory and supergravity generically predict a plethora of scalar fields (moduli or axion-like fields) along with gauge sectors that interact with each other at the non-linear level through dilaton or Chern-Simons like couplings\footnote{The phenomenological roles played by these couplings are initially considered in the context of primordial magnetogenesis \cite{Ratra:1991bn,Garretson:1992vt} and more recently to realize axion-inflation with sub-Planckian decay constants, \eg through strong dissapative dynamics induced by $\rm U(1)$ \cite{Anber:2009ua} or $\rm SU(2)$ \cite{Adshead:2012kp} gauge sectors. For explicit embeddings of the $\rm SU(2)$ model in supergravity and string theory constructions, see \cite{DallAgata:2018ybl} and \cite{McDonough:2018xzh,Holland:2020jdh} respectively.}. In the presence such couplings, the classical roll of the scalar background fields can ``lift" the gauge field fluctuations and enhance their amplitude during inflation in a \emph{parity violating} manner. Produced gauge field modes in this way then can influence the observed tensor fluctuations and can generate a large ``synthetic" component of \emph{chiral} GWs. However, for vector fields that exhibit direct coupling with the observable scalar sector, this is a challenging task because the induced GW emission is also accompanied by the strong production of non-Gaussian scalar fluctuations \cite{Barnaby:2010vf,Barnaby:2011vw,Barnaby:2012tk,Anber:2012du} which puts a bound on the size of the sourced GW component at CMB scales\footnote{At sub-CMB scales however, the same mechanism can be utilized to obtain sufficient enhancement in the scalar fluctuations required for primordial black hole production \cite{Linde:2012bt,Bugaev:2013fya,Erfani:2015rqv,Garcia-Bellido:2016dkw,Domcke:2017fix,Ozsoy:2020kat,Almeida:2020kaq}.} \cite{Ozsoy:2014sba,Mirbabayi:2014jqa}. 

To resolve this tension, an extension of these models are proposed that utilizes spectator axion-$\rm U(1)$ gauge sector endowed with localized gauge field production \cite{Namba:2015gja,Ozsoy:2020ccy}. In this framework, the model is equipped with an additional scalar field that drives inflation and controls observable fluctuations in the scalar sector\footnote{For early studies on the spectator axion-$\rm U(1)$ gauge field model, see \cite{Barnaby:2012xt,Mukohyama:2014gba}, for a discussion on the issues regarding the scalar fluctuations in this model, see \cite{Ferreira:2014zia,Ozsoy:2017blg}. }. Armed with this property and thanks to the localized nature of gauge field production, scalar fluctuations in this model can be kept in observationally viable levels while keeping its original intriguing features such as the generation of chiral GWs of non-vacuum origin. Remarkably, a scan of the parameter space in these model shows that parity violating tensor power spectrum reveals that such a signal may be observable through the mixed angular power spectra\footnote{For earlier studies on probing chiral GWs with CMB anisotropies, see \cite{Lue:1998mq,Saito:2007kt,Gluscevic:2010vv,Ferte:2014gja,Gerbino:2016mqb}.} of the CMB temperature T anisotropies and E,B polarization modes \cite{Namba:2015gja}. 

\begin{figure}[t!]
\begin{center}
\includegraphics[scale=0.4]{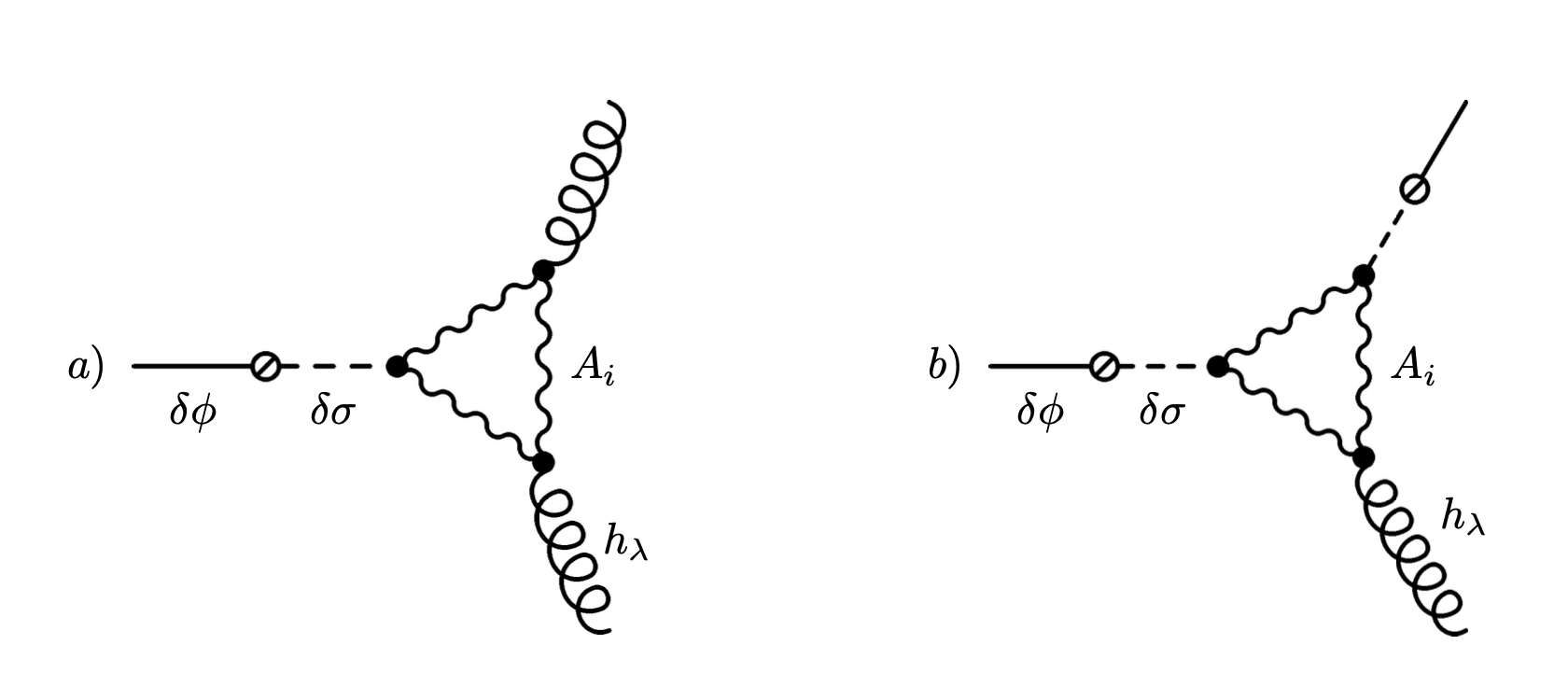}
\end{center}
\caption{Diagrammatic representation of the interactions that contribute to the mixed non-Gaussianity of STT  $\langle \mathcal{R} h h\rangle$ (Left) and TSS $\langle h \mathcal{R} \mathcal{R}\rangle$ (Right) type correlators in the rolling spectator axion-gauge field models. \label{fig:diag}}
\end{figure}

In these models, intriguing parity violating signatures of tensor fluctuations also appear in the tensor-tensor-tensor correlator. In particular, a sizeable, scale dependent tensor non-Gaussianity can be induced by the amplified gauge field fluctuations \cite{Namba:2015gja,Ozsoy:2020ccy} and the parity violation associated with $\langle h_\lambda h_\lambda h_\lambda \rangle$ can reveal itself in the CMB bispectrum of B-modes \cite{Shiraishi:2016yun}. Since scalar fluctuations are also enhanced to a certain extent by the gauge fields, it is then natural to ask if there exist three point cross correlations between scalar and tensor fluctuations. In the spectator axion-$\rm U(1)$ gauge field models, we expect that such mixed correlations to appear on the following grounds: First of all, as we mentioned before, the transient instability in the vector fields can directly influence the metric fluctuations through the inevitable cubic gravitational interaction of $h A A$ type. On the other hand, the scalar fluctuations $\delta \phi$ in the observable inflaton sector can linearly mix with the scalar fluctuations in the spectator axion sector which have direct cubic interactions of $\sgm A A$ type with the gauge field modes. Therefore mediated by the Abelian vector fields, a bridge between the comoving curvature perturbation $\mathcal{R}\propto \delta \phi$ and tensor fluctuations $h$ can be build to induce scalar-tensor-tensor $\mathcal{R}hh$ and tensor-scalar-scalar $h\mathcal{R}\mathcal{R}$ type mixed 3-pt correlators (See Figure \ref{fig:diag}). 

Considering the preferred handedness of tensor fluctuations, along with the scale dependence and the non-gaussian nature of the cosmological fluctuations in these models, a detailed analysis on the mixed bispectra of tensor and scalar perturbations could provide us invaluable information on their underlying production mechanism and guide us compare these predictions with that of the conventional single field models, as well as other non-conventional scenarios\footnote{See \eg \cite{Dimastrogiovanni:2018xnn,Fujita:2018vmv} for an analysis on mixed bispectrum of scalar and tensor fluctuations in the spectator axion-$\rm SU(2)$ gauge field model. On the other hand, parity violating $\langle\mathcal{R}hh\rangle$ bispectrum can also arise through the gravitational Chern-Simons type coupling to the inflation, see \eg \cite{Bartolo:2017szm} and \cite{Bartolo:2018elp} for the detectability of this signal through TBB and EBB CMB bispectra.}. Therefore, for a complete understanding of parity violating signatures in the spectator axion-$\rm U(1)$ gauge field models \cite{Namba:2015gja,Ozsoy:2020ccy}, it is timely to consider 3-pt cross correlations between scalar and tensor fluctuations which is the main focus of this work. 

This paper is organized as follows: in \emph{Section} \ref{S2} we review the transiently rolling spectator axion-gauge field model and its predictions at the level of power and auto bi-spectra. In \emph{Section} \ref{S3}, we present our results on the scalar-tensor-tensor and tensor-scalar-scalar bispectrum and discuss their amplitude and shape dependence. We conclude in \emph{Section} \ref{S4}. We supplement our results with five appendices where many details about the computations we carry can be found.

{\bf Notations and conventions.}  Our metric signature is mostly plus sign $(-,+,+,+)$. Greek indices stand for
space-time coordinates, while Latin indices denote spatial coordinates. Overdots and primes on time dependent quantities will denote derivatives with respect to coordinate time $t$ and conformal time $\tau$, respectively. At leading order in slow-roll parameters, we take the scale factor as $a(\tau) = 1/(-H \tau )$ where $H = \dot{a}/a$ is the physical Hubble rate during inflation.

\section{Cosmological fluctuations from axion-gauge field dynamics}\label{S2}
As we mentioned in the introduction, the Lagrangian that describes the model contains an inflationary sector together with a spectator axion-gauge field sector both minimally coupled to gravity  \cite{Barnaby:2012xt,Namba:2015gja,Peloso:2016gqs,Ozsoy:2020ccy},
\beq\label{Lm}
\fr{\mathcal{L}}{\sqrt{-g}} =\fr{\Mp^2 R}{2}+ \mathcal{L}_\phi-\underbrace{\fr{1}{2}(\del\sigma)^2 - V_{\sgm}(\sgm)-\fr{1}{4}F_{\mu\nu}F^{\mu\nu}-\fr{\alpha_{\rm c}\sgm}{4f}F_{\mu\nu}\tilde{F}^{\mu\nu},}_\text{Spectator Sector}
\eeq
where $\mathcal{L}_\phi$ is the Lagrangian that drives inflation and is responsible for the generation of curvature perturbation consistent with CMB observations and $\sgm$ is a spectator pseudo-scalar axion rolling on its potential $V_\sgm$. The strength of the interaction (\ie the last term in \eqref{Lm}) between the spectator $\sgm$ and the gauge field is parametrized by the scale $f$ together with the dimensionless coupling constant $\alpha_c$ where $F_{\mu\nu} = \del_\mu A_\nu - \del_\nu A_\mu$  is the field strength tensor of $\rm U(1)$ gauge field, $\tilde{F}^{\mu\nu}\equiv \eta^{\mu\nu\rho\sigma} F_{\rho\sigma}/(2\sqrt{-g})$ is its dual and alternating symbol $\eta^{\mu\nu\rho\sigma}$ is $1$ for even permutation of its indices, $-1$ for odd permutations, and zero otherwise.

\smallskip
\noindent
{\bf Gauge field production.} If the spectator axion rolls on its potential $V_\sgm$ with a non-vanishing background velocity $\dot{\sgm} \neq 0$, the interaction in \eqref{Lm} introduces a tachyonic mass for the gauge field and leads to the enhancement of gauge field modes in a parity violating manner. This can be seen from the equation of motion of the gauge field polarization states $A_\pm$ in a FRW background \cite{Anber:2009ua},
\beq\label{nmea}
\left(\partial_x^2 + 1 \pm \fr{aH}{k}2\xi\right) A_{\pm} = 0,
\eeq
where we defined $x \equiv -k\tau$ and $\xi \equiv - \alpha_{\rm c} \dot{\sgm}/ (2Hf)$ ($\xi > 0$ \& $\dot{\sgm} < 0$) is the dimensionless measure of axion's velocity that represents the effective coupling strength between $\sgm$ and $A_\mu$.  From \eqref{nmea},  we see that when the last term dominates over unity for $k/(aH) < 2\xi$,  only the $-$ polarization state of the gauge field experiences tachyonic instability which reflects the parity-violating nature of the $\sgm F\tilde{F}$ interaction. 

\smallskip
\noindent{\bf Tensors sourced by vector fields.} The gauge field fluctuations produced in this way exhibit an amplitude $A_{-} \propto e^{\pi \xi}$ \cite{Anber:2009ua} which in turn act as an additional source of tensor perturbations through gravitational interactions \cite{Barnaby:2012xt}. This can be seen clearly from the mode equation of graviton polarization states $h_{\lambda}$ which is sourced by the transverse, traceless part of the energy momentum tensor composed of gauge field fluctuations:
\beq\label{te}
\left(\partial^2_\tau + k^2 -\fr{2}{\tau^2}\right)(a \hat{h}_\lambda) =-\frac{2a^{3}}{\Mp^2} \Pi_{i j, \lambda}(\vec{k}) \int \frac{\mathrm{d}^{3} p}{(2 \pi)^{3 / 2}}\left[\hat{E}_{i}(\tau, \vec{k}-\vec{p}) \hat{E}_{j}(\tau,\vec{p})+\hat{B}_{i}(\tau,\vec{k}-\vec{p}) \hat{B}_{j}(\tau,\vec{p})\right],
\eeq
where $\hat{E}_i = -{a^{-2}} \hat{A}_{i}^{\prime}, \quad \hat{B}_{i}={a^{-2}} \epsilon_{i j k} \partial_{j} \hat{A}_{k}$ are ``electric" and ``magnetic" fields and $\hat{h}_\lambda (\tau, \vec{k}) = \Pi_{ij,\lambda} (\vec{k})\,\hat{h}_{ij}(\tau,\vec{k})$ with $\Pi_{ij,\lambda}$ being the polarization tensor obeying $\hat{k}_i\, \Pi_{ij,\lambda}(\vec{k})=0 $,  $\Pi^{*}_{ij,\lambda}\Pi_{ij,\lambda'} = \delta_{\lambda\lambda'}$ and $\Pi^{*}_{ij,\lambda}(\vec{k}) = \Pi_{ij,-\lambda}(\vec{k}) = \Pi_{ij,\lambda}(-\vec{k})$.

\smallskip
\noindent{\bf Scalars sourced by vector fields.} The influence of particle production on the visible scalar sector is also encoded indirectly by the presence of gravitational interactions \cite{Ferreira:2014zia}. In particular, integrating out the non-dynamical lapse $\delta N$ and the shift $N^{i}$ reveals a mass mixing between $\delta \phi$ and $\delta \sgm$ and opens up a channel that can influence the curvature perturbation\footnote{In the multi-field model we are considering, late time $\mathcal{R}$ also obtains direct contributions from fluctuations linear in the spectator axion $\delta \sgm$ and the gauge fields at non-linear order. For a spectator axion that rolls down to its minimum long before the end of inflation -- as we assume in this work -- the contribution of $\delta \sgm$ can be neglected \cite{Mukohyama:2014gba,Namba:2015gja}. The contribution from gauge fields on the other hand is roughly proportional to the absolute value of Poynting vector, $a|\vec{S}| = a|\vec{E}\times \vec{B}|$ which is also negligible at late times as the particle production saturates at super-horizon scales and the resulting electromagnetic fields decay as $\vec{E},\vec{B} \sim a^{-2}$ \cite{Ozsoy:2020ccy}. For the purpose of evaluating mixed correlators, we therefore adopt the standard relation $\mathcal{R} \equiv  - H \delta \phi/\dot{\phi}$ in this work.} $\mathcal{R} \simeq -{H\,\delta \phi}/{\dot{\phi}} $ through the inverse decay of gauge fields: $A_i + A_i \to \delta \sgm \to \delta \phi \propto \mathcal{R}$. Dynamics of this contribution can be understood by first studying the influence of particle production on the spectator fluctuations $\delta \sgm$ through,
\beq
\label{usgm}\left(\frac{\partial^{2}}{\partial \tau^{2}}+k^{2}-\frac{2}{\tau^{2}}\right) (a \delta \hat{\sgm}) \simeq a^3\fr{\alpha_{\rm c}}{f}\int\fr{\d^3 p}{(2\pi)^{3/2}}~ \hat{E}_{i}(\tau, \vec{k}-\vec{p}) ~\hat{B}_{i}(\tau,\vec{p}).
\eeq
Focusing on the inhomogeneous solution of the $\delta \sgm$ fluctuations in \eqref{usgm}, one can then compute the conversion of the resulting $\delta \sgm$ to $\delta \phi$ via 
\beq
\label{uphi} \left(\frac{\partial^{2}}{\partial \tau^{2}}+k^{2}-\frac{2}{\tau^{2}}\right) (a \delta \hat{\phi}) \simeq 3a^2 \fr{\dot{\phi}\dot{\sgm}}{\Mp^2} ~(a \delta \hat{\sgm}),
\eeq
to find the the part of curvature perturbation that is sourced by the amplified gauge fields.

It has recently been shown that if $\sgm$ rolls for a large-amount of time ($\Delta N_\sgm \gg 1$) during inflation, the sourced contributions to the $\mathcal{R}$ can be sizeable due to the sensitivity of gauge field amplitudes and $\delta\phi-\delta\sgm$ mixing on the spectator axion's velocity $\xi \propto |\dot{\sgm}|$ \cite{Ferreira:2014zia}. In particular, this would lead to an exceedingly large CMB non-Gaussianity and once the CMB limits on it are respected, the sourced GW signal is bounded by $r < 10^{-3}-10^{-4}$ at CMB scales \cite{Ferreira:2014zia,Ozsoy:2017blg}. To minimize the influence of the enhanced gauge fields on the curvature perturbation and to render observable GWs sourced by gauge fields viable, more realistic models that lead to localized gauge field production has been proposed where the spectator axion transiently rolls on potentials of the following form \cite{Namba:2015gja,Ozsoy:2020ccy}: 
\beq\label{pots}
V_\sgm(\sgm)=
 \begin{dcases} 
       \Lambda^4 \left[1-\cos\left(\fr{\sgm}{f}\right)\right],& \quad{\rm Model\, 1}\,({\rm M}1) \,,\\
        \mu^3\sgm + \Lambda^4 \left[1-\cos\left(\fr{\sgm}{f}\right)\right]\&\,\, \Lambda^4\lesssim \mu^3 f &\quad{\rm Model\, 2}\,({\rm M}2).
   \end{dcases}
\eeq 

The first model (M1) features a spectator axion with standard shift symmetric potential (see \eg \cite{Freese:1990rb}) where the size of the axion modulations is set by the mass parameter $\Lambda$. In this model, the motion of the axion is contained within the maximum ($\sgm = \pi f$) and the minimum ($\sgm = 0$) of the potential whereas in the second model (M2), the axion field range is extended via a monodromy term \cite{McAllister:2008hb,McAllister:2014mpa} proportional to a second mass parameter $\mu$ and $\sgm$ is assumed to probe step-like feature(s) in the ``bumpy" regime, $\Lambda^{4}\lesssim \mu^3 f$ \footnote{In the bumpy regime, depending on the initial conditions ($\sgm \gg f$) spectator axion can probe multiple step-like features during inflation. In this work, we assume that $\sgm$ traverse only one such region on its potential during which observable scales associated with CMB exits the horizon.}.

For the typical field ranges dictated by the scalar potentials \eqref{pots} and assuming slow-roll $\ddot{\sgm} \ll 3H\dot{\sgm}$ condition, the spectator field velocity $\dot{\sgm}$ and the effective coupling $\xi = -\alpha_{\rm c} /(\dot{\sgm}/2Hf)$ in \eqref{nmea} obtains a peaked time dependent profile given by \cite{Namba:2015gja,Ozsoy:2020ccy},
\beq\label{Joep}
\xi(\tau)=
 \begin{dcases} 
       \frac{2 \xi_{*}}{\left({\tau_*}/{\tau}\right)^{\delta}+\left({\tau}/{\tau_*}\right)^{\delta}},& \quad{\rm Model\, 1}\,({\rm M}1) \,,\\
        \frac{\xi_{*}}{1+\ln[\left({\tau}/{\tau_*}\right)^{\delta}]^2},&\quad{\rm Model\, 2}\,({\rm M}2),
   \end{dcases}
\eeq 
where $\xi_* = \{\alpha_{\rm c} \delta/2\,\, (\rm M1), \alpha_{\rm c} \delta\,\, (\rm M2)\}$ is the maximum value of $\xi$ when the axion's velocity becomes maximal  at the conformal time $\tau_*$. In \eqref{Joep}, we defined the dimensionless ratios $\delta = \Lambda^4 / 6H^2 f^2$ (M1) and $\delta \simeq \mu^3 / 3H^2 f$ (M2) in terms of the model parameters. Physically, $\delta$ is a measure for the acceleration ($\dot{\xi}/{(\xi H)} = \ddot{\sgm}/(\dot{\sgm}H) \sim \delta $) of the spectator axion as it rolls down on its potential. Note that since the slow-roll approximation $\ddot{\sgm} \ll 3H\dot{\sgm}$ is assumed to derive \eqref{Joep}, we require $\delta < 1$. In this work, without loss of generality we will adopt $\delta = 0.3$ \footnote{We note that this choice is not a unique requirement for successful phenomenology and other values for $\delta$ can be adopted (see \eg \cite{Namba:2015gja}) as far as we restrict ourselves to $0 \leq \delta < 1$. However, different choices of $\delta$ within this range influence the properties of the scale dependent signals as we explain in section \ref{S3}. For example, $\delta \to 0$ limit corresponds to the standard scale invariant production of gauge fields (with a constant $\xi$) for an axion rolling at a constant rate \cite{Anber:2009ua,Barnaby:2010vf}. We refer the reader to \cite{Namba:2015gja,Ozsoy:2020ccy} for many details regarding the parameter $\delta$ including its relation with axion dynamics, particle production in the gauge field sector and the resulting phenomenology of scalar and tensor correlators.} to derive phenomenological implications of spectator axion-gauge field dynamics. 

As we review in Appendix \ref{AppA}, the time dependent profile \eqref{Joep} for $\xi$ translates into a scale dependent growth of the gauge fields in \eqref{nmea} where only modes that has a size comparable to the horizon, \ie $k \simeq \mathcal{O}(1)\, a_*H_*$ at $\tau= \tau_*$, are efficiently amplified. Below we review the impact of such scale dependent vector field production on the auto correlators of tensor and scalar fluctuations during inflation. 
\begin{figure}[t!]
\begin{center}
\includegraphics[scale=0.89]{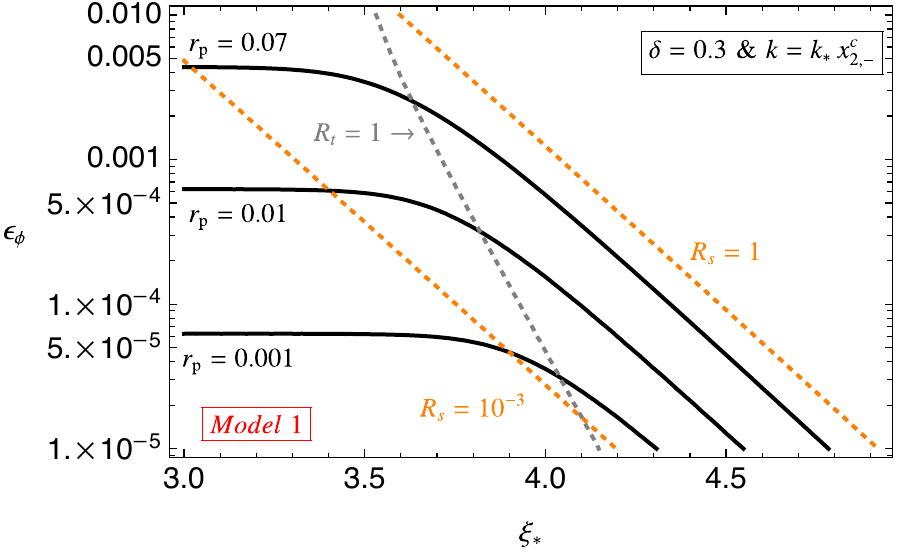}\includegraphics[scale=0.89]{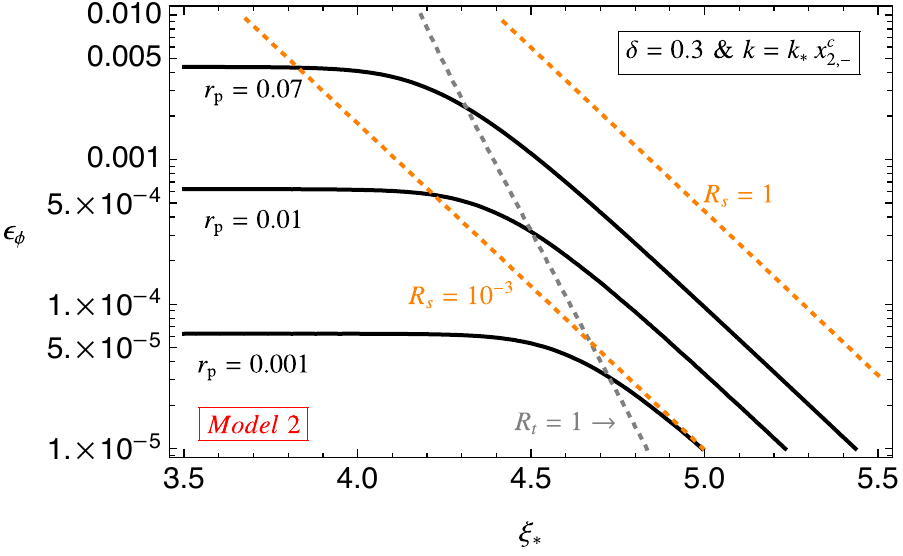}
\end{center}
\caption{Constant $r$ curves in the $\epsilon_\phi -\xi_*$ plane for Model 1 (Left) and Model 2 (Right). Orange dotted (respectively, gray dotted) lines shows the ratio between the sourced and the vacuum scalar (respectively, tensor) power spectrum $R_s$ ($R_t$) (See Appendix \ref{AppB} for details).\label{fig:rpeak}}
\end{figure}

\smallskip
\noindent{\bf Chiral GWs from gauge field sources.} In the presence of gauge field amplification, the perturbations in the observable sector $\hat{\mathcal{X}} = \{\hat{\mathcal{R}},\hat{h}_\pm\}$ pick up a sourced contribution that can be described by the particular solutions of \eqref{te} and \eqref{uphi} (see also \eqref{usgm}) in addition to the vacuum counterpart generated by quasi-dS background: $\hat{\mathcal{X}} = \hat{\mathcal{X}}^{(\rm v)} + \hat{\mathcal{X}}^{(\rm s)}$. These contributions are statistically uncorrelated and therefore the total power spectra can be simply described by the sum of vacuum and sourced part: 
\beq\label{tps}
\mathcal{P}_{\mathcal{R}}(k) = \mathcal{P}^{(\rm v)}_{\mathcal{R}}(k) + \mathcal{P}^{(\rm s)}_{\mathcal{R}}(k),\quad\quad \mathcal{P}_{\pm}(k) = \mathcal{P}^{(\rm v)}_{\pm}(k) + \mathcal{P}^{(\rm s)}_{\pm}(k),
\eeq
where the vacuum contributions are given by the standard expressions:
\beq\label{psv}
\mathcal{P}_{\mathcal{R}}^{(\rm v)} = \fr{H^2}{8\pi^2\epsilon_\phi \Mp^2},\quad\quad \mathcal{P}^{(\rm v)}_{\pm} = \fr{H^2}{\pi^2 \Mp^2},
\eeq
with $\epsilon_\phi \equiv \dot{\phi}^2/ (2 H^2 \Mp^2)$ is the slow-roll parameter controlled by the inflaton sector. 
On the other hand, the sourced power spectra in \eqref{tps} inherit the scale dependence of the gauge field sources which can be shown to acquire a Gaussian form \cite{Namba:2015gja,Ozsoy:2020ccy},
\begin{align}\label{SC}
\nn \mathcal{P}^{(\rm s)}_{j}(k) &=\left[\epsilon_{\phi} \mathcal{P}_{\mathcal{R}}^{(\rm v)}(k)\right]^{2} f_{2, j}\left(\xi_{*},\frac{k}{k_{*}}, \delta\right), \\
f_{2, j}\left( \xi_{*},\frac{k}{k_{*}}, \delta\right) & \simeq f_{2, j}^{c}\left[\xi_{*}, \delta\right] \exp \left[-\frac{1}{2 \sigma_{2, j}^{2}\left[\xi_{*}, \delta\right]} \ln ^{2}\left(\frac{k}{k_{*} x_{2, j}^{c}\left[\xi_{*}, \delta\right]}\right)\right],
\end{align}
where $j = \{\mathcal{R}, \pm\}$. The functions $ f_{2, j}^{c}, \sigma_{2, j}, x_{2, j}^{c}$ control, respectively, the amplitude, the width, and the position of the peak of the sourced signal, which depend on the background model of the spectator axion through the parameters $\xi_*$ and $\delta$ we discussed above and therefore to the underlying scalar potential \eqref{pots} in the spectator axion sector. For a representative choice of the background parameter $\delta$, we present accurate formulas for $ f_{2, j}^{c}, \sigma_{2, j}, x_{2, j}^{c}$ in terms of the effective coupling $\xi_*$ in Table \ref{tab:fit1}.  

At this point, it is intriguing to ask if the gauge field sources can be sufficiently large to alter tensor-to-scalar ratio defined by  \cite{Namba:2015gja,Ozsoy:2020ccy},
 
 \beq\label{defr}
 r (k)= \sum_\lambda \fr{\mathcal{P}^{(\rm v)}_\lambda (k) + \mathcal{P}^{(\rm s)}_\lambda(k)}{\mathcal{P}^{(\rm v)}_{\mathcal{R}}(k)+\mathcal{P}^{(\rm s)}_{\mathcal{R}}(k)}.
 \eeq
 \noindent To address this question, in Figure \ref{fig:rpeak} we show constant curves of tensor-to-scalar ratio $r$ \eqref{defr} evaluated at the peak of the sourced GW signal $r (k_{\rm p} = k_*\,x^c_{2,-}) = r_p$, in the $\epsilon_\phi - \xi_*$ plane for both models (see Appendix \ref{AppB}). In this plot, the region spanned between the $R_t \equiv \mathcal{P}_{-}^{(\rm s)}/{\mathcal{P}_{h}^{(\rm v)}}>1$ and $R_s \equiv  \mathcal{P}_{\mathcal{R}}^{(\rm s)}/{\mathcal{P}_{\mathcal{R}}^{(\rm v)}} \ll 1$ locates the parameter space where sourced GWs dominate over the vacuum fluctuations while keeping the amplitude of scalar fluctuations sourced by the gauge fields are under control \footnote{In the $R_s \ll 1$ regime, additional limitations on the model parameter space arise from the CMB constraints on the spectral tilt and its running. For both models ({\rm M1},{\rm M2}) we consider, a detailed discussion on these limitations appeared in \cite{Peloso:2016gqs,Ozsoy:2020ccy} where it was found that axion decay constants that roughly obeys $f/\Mp < 0.1$ (at fixed $\delta$) are preferred in order to grant observable GWs of non-vacuum origin. Note that this bound does not lead to an additional constraint on the amplitude of the signals sourced by the gauge fields as the latter mainly controlled by $\xi_*$ or equivalently by the dimensionless coupling constant $\alpha_{\rm c}$ at fixed $\delta$, considering the relation $\xi_* \propto \alpha_{\rm c} \delta$. }. In this region, tensor-to-scalar ratio $r$ acquires an exponential sensitivity to gauge field production (See \eqref{rpeak}), breaking the standard relation between $r$ and $H$ of single field inflation. Excitingly, GW signal produced by the gauge field sources is maximally chiral $\chi \equiv (\mathcal{P}_{-} - \mathcal{P}_{+})/\sum_\lambda \mathcal{P}_\lambda \sim \mathcal{O}(1)$ (see \eg \cite{Sorbo:2011rz}) which is an essential distinguishing feature of the inflationary models we consider in this work\footnote{In contrast to standard predictions of inflation, chiral GWs can produce a non-vanishing cross correlation between CMB temperature (T) anisotropies and polarization modes (E,B) \cite{Lue:1998mq,Gluscevic:2010vv,Saito:2007kt}. See \eg \cite{Namba:2015gja}, for an analysis on the observability of the CMB TB correlator within the first model we present here. On the other hand, the observability of a chiral GW signal in the spectator axion-SU(2) gauge field model is studied in \cite{Thorne:2017jft}.}. 

\smallskip
\noindent{\bf Scalar and tensor bispectrum.} The scale dependent amplification of gauge fields also influences 3-pt correlators of scalar and tensor perturbations. An immediate worry at this point is to keep scalar bispectrum $\langle\hat{\mathcal{R}}_{k_1}\hat{\mathcal{R}}_{k_2}\hat{\mathcal{R}}_{k_3} \rangle$ below the CMB observational limits while preserving a large chiral GW signal from gauge field sources. This issue is addressed in \cite{Namba:2015gja,Ozsoy:2020ccy} for both spectator axion-gauge field models where it was shown that stringent constraints on scalar non-Gaussianity at CMB scales can be avoided for much of the parameter space of these models, thanks to the localized nature of particle production in the gauge field sources. Remarkably, a sizeable parity violating tensor non-Gaussianity  $\langle\hat{h}_{k_1}\hat{h}_{k_2}\hat{h}_{k_3} \rangle$ \footnote{Observably large tensor non-Gaussianity can also arise from spectator axion-${\rm SU}(2)$ gauge field dynamics during inflation \cite{Agrawal:2017awz,Agrawal:2018mrg}.} can also be generated by the gauge field sources, providing an opportunity to test these models through the CMB B-mode bispectrum \cite{Shiraishi:2016yun}. 

For the distinguishability of these signals, shape dependence 3-pt auto-correlators will provide further information. The shape analysis is carried for the first model discussed (See M1 in \eqref{pots}) \cite{Namba:2015gja} , where it was shown that both bispectrum is maximal at the equilateral configurations, $k_1 \approx k_2 \approx k_3 \simeq \mathcal{O}(1-10)\,k_*$. In Appendix \ref{AppBB}, we likewise perform the shape analysis of the bispectra for the non-compact axion model (M2) in \eqref{pots} to confirm that both scalar and tensor bispectrum is also maximal at the equilateral configuration in this model. The appearance of the equilateral shape in the auto-correlators is closely tied to the gauge field sources which have maximal support only for modes satisfying $q \simeq \mathcal{O}(1)\, a_* H_*$ (See Table \ref{tab:fit0}).

Due to scale dependent amplifications of tensor (T) and scalar (S) fluctuations, their 3-pt cross correlations may also contain invaluable information on the production mechanism of primordial GWs and more importantly on the inflationary field content. Considering chirality of the tensor fluctuations present in these models, the size and the shape of the mixed non-Gaussianity is complementary to the auto-correlators of $\hat{\mathcal{R}}$ and $\hat{h}$ in extracting this unique information and can help us distinguish this class of models from other scenarios. In what follows, we will study the mixed non-Gaussianity of scalar-tensor-tensor $\langle\hat{\mathcal{R}}\hat{h}\hat{h}\rangle$ (STT) and tensor-scalar-scalar $\langle \hat{h}\hat{\mathcal{R}}\hat{\mathcal{R}}\rangle$ (TSS) type during inflation focusing on the spectator axion-gauge field dynamics described by the potentials \eqref{pots}.

\section{Mixed non-Gaussianity from axion-gauge field dynamics}\label{S3}
In the theory described by the Lagrangian \eqref{Lm}, the last two terms contain three legged vertices $\propto \delta \sgm F\tilde{F}$ and $\propto h_{ij} \{\dot{A}_i\dot{A}_j + \dots \}$ that capture the inverse decay of amplified gauge fields fluctuations to the spectator scalar and tensor fluctuations, respectively \cite{Barnaby:2012xt,Ozsoy:2017blg}. The presence of these vertices ensure correlations between the observable scalar sector $\delta \phi \propto \mathcal{R}$ and the metric $h_{ij}$ perturbations thanks to the mass mixing between $\delta \phi-\delta \sgm$ we discussed earlier. Therefore, we expect the effects of the particle production processes in the gauge field sector to propagate to the 3-pt functions of mixed type such as $\langle \hat{\mathcal{R}} \hat{h} \hat{h}\rangle$ \footnote{There is an additional four legged vertex $hhAA$ that appear at the same order as the three legged vertex $h A A$ (see Figure \ref{fig:diag}) in the gravitational coupling $\Mp^{-2}$ \cite{Eccles:2015ipa}. Combined with a three legged scalar vertex $\delta \sigma A A$ and $\delta \phi-\delta \sgm$ mixing, $hhAA$ leads to an additional diagram that contributes to STT correlator. However, such a diagram contains fewer internal gauge field modes compared to the left diagram in Figure \ref{fig:diag} and thus carry less particle production effects. In particular, counting the number of gauge field modes that contributes to the loop integral (see \eg \eqref{f3RMM}), we anticipate that the diagram that includes $hhAA$  will be suppressed by a factor of $e^{-2c\pi\xi_*}$ ($c \simeq \mathcal{O}(1)$, see Table \ref{tab:fit0}) compared to diagram we are computing in this work.} and $\langle \hat{h} \hat{\mathcal{R}} \hat{\mathcal{R}}\rangle$. The diagrams that contribute to these non-Gaussianities can be pictorially represented as in Figure \ref{fig:diag}. In Appendix \ref{AppC}, we calculate both types of mixed non-Gaussianity for the two different rolling axion spectator models (See eq. \eqref{pots}) we introduced in the previous section. In the following sections we present our results and discuss their size and shape dependence.
\subsection{Results for TSS and STT type correlators}
We are interested in 3-pt cross correlation of comoving curvature perturbation and gravity wave polarization modes, in particular in the following mixed type non-Gaussian correlators that are defined by
\begin{align}\label{DBS}
\nn \left\langle\hat{\mathcal{R}}\left(0, \vec{k}_{1}\right) \hat{h}_\lambda\left(0, \vec{k}_{2}\right) \hat{h}_\lambda\left(0, \vec{k}_{3}\right)\right\rangle &\equiv \mathcal{B}_{\mathcal{R}\lambda\lambda}\left(\vec{k}_{1}, \vec{k}_{2}, \vec{k}_{3}\right) \delta\left(\vec{k}_{1}+\vec{k}_{2}+\vec{k}_{3}\right),\\
\left\langle \hat{h}_\lambda\left(0, \vec{k}_{1}\right)\hat{\mathcal{R}}\left(0, \vec{k}_{2}\right)\hat{\mathcal{R}}\left(0, \vec{k}_{3}\right)\right\rangle &\equiv \mathcal{B}_{\lambda\mathcal{R}\mathcal{R}}\left(\vec{k}_{1}, \vec{k}_{2}, \vec{k}_{3}\right) \delta\left(\vec{k}_{1}+\vec{k}_{2}+\vec{k}_{3}\right).
\end{align}
As we mentioned in the case of 2-pt correlators above, mixed type non-Gaussianities are given by simple sum of vacuum and sourced contributions:  $\mathcal{B}_{j} = \mathcal{B}^{(\rm v)}_{j} + \mathcal{B}^{(\rm s)}_{j} $. In this work, we will disregard the vacuum component of mixed correlators as they are sub-dominant in the presence of particle production processes involving vector fields.  

As in the case of 2-pt functions, mixed 3-pt correlators of $\hat{\mathcal{R}}^{(\rm s)}$, $\hat{h}^{(\rm s)}_{-}$ inherit the scale dependent amplification of vector fields triggered by the transient motion of the spectator axion $\sgm(t)$. In particular, we found (See Appendix \ref{AppC}) that both bispectra can be factorized as
\beq\label{SC3pt}
\mathcal{B}^{(\rm s)}_{\,\,\,j}(\vec{k}_1,\vec{k}_2,\vec{k}_3) =  \frac{\left[\epsilon_{\phi} \mathcal{P}_{\mathcal{R}}^{(\rm v)}\right]^{3}}{(k_{1} k_{2} k_{3})^2}\, f^{(3)}_{\,\,\,j}\left(\xi_{*}, \delta, x_{*}, x_{2}, x_{3}\right)\\
\eeq
where $f^{(3)}_j$ with $j = \{\mathcal{R}\lambda\lambda, \lambda\mathcal{R}\mathcal{R}\}$ are dimensionless functions that parametrize the scale and shape dependence of the bispectrum noting the definitions $x_* = k/k_*$ and $k_1 = k,\, x_2 = k_2/k_1,\,x_3 = k_3/k_1$. In the following, we will first focus on the scale dependence of the mixed bispectra to set the stage for a discussion on its amplitude and shape dependence. In our analysis, we found some qualitative differences between TSS and STT type mixed 3-pt correlators and hence we will discuss each case separately below. 

\subsubsection{TSS correlators}\label{S3p1p1}
To study the scale dependence of $f^{(3)}_{\lambda\mathcal{R}\mathcal{R}}$, we focus on the equilateral configuration to work out the $x_* = k/k_*$ dependence of \eqref{f3MRR} at fixed values of the background parameters $\{\xi_*, \delta \}$. In the following we will discuss $-\mathcal{R}\mathcal{R}$ and $\{+\mathcal{R}\mathcal{R}\}$ type correlators separately.
\smallskip

\begin{figure}[t!]
\begin{center}
\includegraphics[scale=0.89]{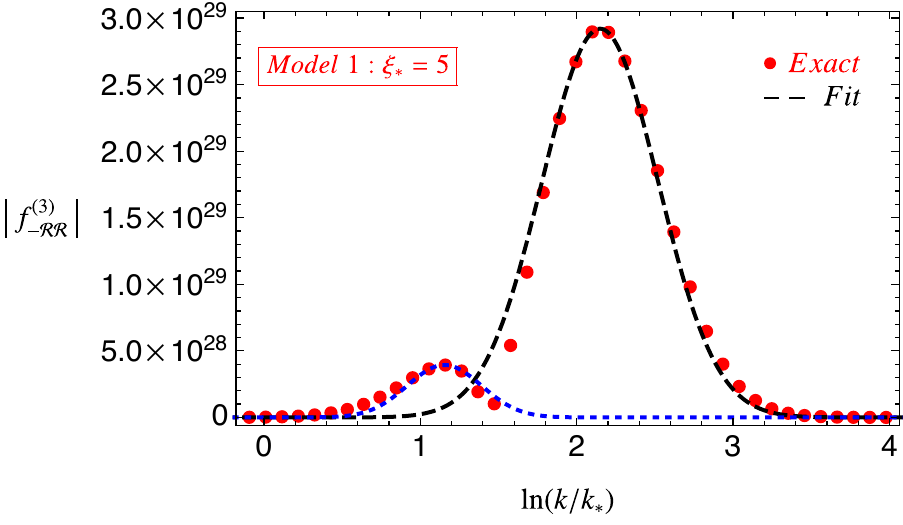}\includegraphics[scale=0.89]{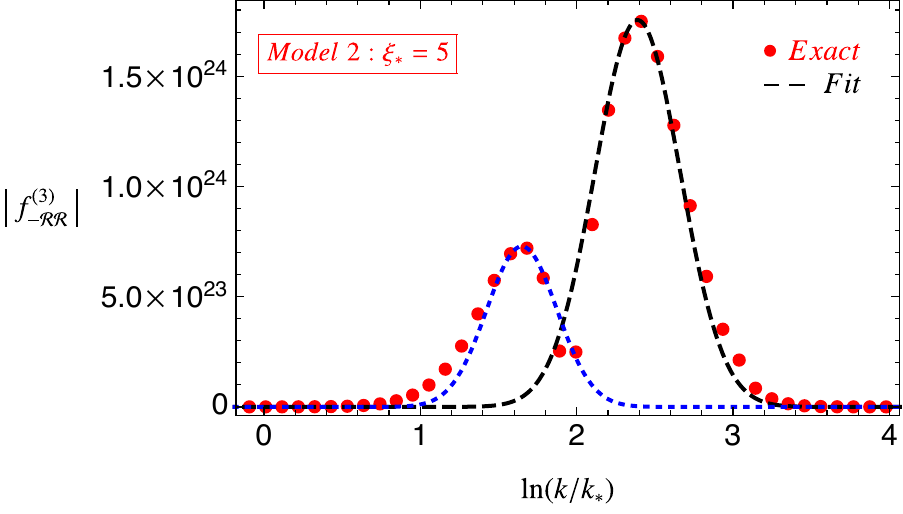}
\end{center}
\caption{The scale dependence of total $f^{(3)}_{-\mathcal{R}\mathcal{R}}$ for Model 1 (Left) and Model 2 (Right). The red points are obtained by direct numerical evaluation of \eqref{f3MRR} for a grid of $x_* = k/k_*$ values. Using the Table \ref{tab:f3fit}, we represent the accuracy of the Gaussian expression \eqref{f3fit} (dashed lines) in parametrizing the large and small peak that constitutes the total signal. Note that the first peak occurs while  $f^{(3)}_{-\mathcal{R}\mathcal{R}} >0$ while for the second peak $f^{(3)}_{-\mathcal{R}\mathcal{R}} < 0$. \label{fig:f3fit}}
\end{figure}

{\bf $\bullet\,\, -\mathcal{R}\mathcal{R}$ bispectrum:} Computing \eqref{f3MRR} numerically for a grid of $x_* = k/k_*$ values at the equilateral configuation $x_2 = x_3 =1$, we found that $-\mathcal{R}\mathcal{R}$ bispectrum can be accurately captured by a sum of two distinctive peaks $f^{(3)}_{-\mathcal{R}\mathcal{R}} = f^{(3,S)}_{-\mathcal{R}\mathcal{R}} + f^{(3,L)}_{-\mathcal{R}\mathcal{R}}$ that have the Gaussian form
\beq\label{f3fit}
f^{(3,\alpha)}_{j}\left( \xi_{*},\frac{k}{k_{*}}, \delta\right)  \simeq f_{3,j}^{c,\alpha}\left[\xi_{*}, \delta\right] \exp \left[-\frac{1}{2 \sigma^{\alpha}_{3, j}\left[\xi_{*}, \delta\right]^2} \ln ^{2}\left(\frac{k}{k_{*} x_{3, j}^{c,\alpha}\left[\xi_{*}, \delta\right]}\right)\right],
\eeq
where we use $\alpha =\{S,L\}$ to label the each peak (\ie a small and a large one) and $j =-\mathcal{R}\mathcal{R}$. As in the 2-pt correlators we mentioned earlier, the height $f_{3,j}^{c}$, width $\sigma_{3, j}$ and location $x_{3, j}^{c}$ of $f^{(3)}_{ j}$'s peak is controlled by the background motion of the spectator axion, namely by the maximal velocity reaches $\xi_* = -\alpha_{\rm c} \dot{\sgm}_{*} / 2Hf$ and the total number of e-folds $\dot{\sgm}$ significantly differs from zero during its rollover: $\Delta N \sim \delta^{-1} \sim H^2/m_{\rm axion}^2$ where $m_{\rm axion}$ is the mass of $\sgm$ in its global minimum. At fixed value of $\xi_*$, increasing $\delta$, would generically reduce $f_{3,j}^{c}$, width $\sigma_{3, j}$ and location $x_{3, j}^{c}$ because fewer gauge field modes can be amplified to excite cosmological perturbations as $\dot{\sgm}$ will be large for a shorter amount of time in this case. For $\delta=0.3$, we determined $\xi_*$ dependence of $f_{3,j}^{c}$, $\sigma_{3, j}$ and $x_{3, j}^{c}$ by fitting the right hand side of eq. \eqref{f3fit} to reproduce the position, height and width of the sourced peaks parametrized by the integral \eqref{f3MRR}. In Table \ref{tab:f3fit}, we present the  $\xi_{*}$ dependence of these fitting formulas appear in \eqref{f3fit} that approximates the result from the direct numerical integration of \eqref{f3MRR}. For both models we study in this work, the accuracy of the expression \eqref{f3fit} is shown in Figure \ref{fig:f3fit}.

\begin{table}[t!]
\begin{center}
\begin{tabular}{|c|c|c|c|}
\hline
\hline
\cellcolor[gray]{0.9}$\{j\}_{\alpha}$&\cellcolor[gray]{0.9}$\ln(|f^c_{3,j}|)$&\cellcolor[gray]{0.9}$x^c_{3,j}$&\cellcolor[gray]{0.9}$\sgm_{3,j}$ \\
\hline
\cellcolor[gray]{0.9}\scalebox{0.9}{$\{-\mathcal{R}\mathcal{R}\}_{\rm S,M1}$}&\scalebox{0.9}{$-20.75 + 17.87\,\xi_*- 0.109\,\xi_*^2$}&\scalebox{0.9}{$-2.59 +1.336\, \xi_* - 0.0366\, \xi_* ^2$}&\scalebox{0.9}{$0.48 -0.166\, \xi_* + 0.0234\, \xi_* ^2$}\\\hline
\cellcolor[gray]{0.9}\scalebox{0.9}{$\{-\mathcal{R}\mathcal{R}\}_{\rm L,M1}$} &\scalebox{0.9}{$-8.20 +14.60\, \xi_* + 0.121\, \xi_* ^2$}&\scalebox{0.9}{$2.94 + 0.980\,\xi_* + 0.0294\, \xi_*^2$}&\scalebox{0.9}{$0.85 -0.133\,\xi_* + 0.0076\,\xi_* ^2 $}\\\hline
\cellcolor[gray]{0.9}\scalebox{0.9}{$\{+\mathcal{R}\mathcal{R}\}_{\rm M1}$} &\scalebox{0.9}{$-10.10 +14.69\, \xi_* + 0.119\, \xi_* ^2$}&\scalebox{0.9}{$2.66 + 0.516\,\xi_* + 0.0195\, \xi_*^2$}&\scalebox{0.9}{$0.83 -0.108\,\xi_* + 0.0065\,\xi_* ^2 $}\\\hline
\hline
\hline
\cellcolor[gray]{0.9}\scalebox{0.9}{$\{-\mathcal{R}\mathcal{R}\}_{\rm S,M2}$}&\scalebox{0.9}{$ -29.67 + 17.53\,\xi_*- 0.121\,\xi_*^2$} & \scalebox{0.9}{$-0.78 + 1.345\, \xi_* - 0.0295\, \xi_* ^2$}&\scalebox{0.9}{$\,\,-0.34 + 0.224\,\xi_* - 0.0222\, \xi_* ^2$ }\\\hline
\cellcolor[gray]{0.9}\scalebox{0.9}{$\{-\mathcal{R}\mathcal{R}\}_{\rm L,M2}$}&\scalebox{0.9}{$-20.34 +14.88\, \xi_* + 0.0708\, \xi_* ^2$}&\scalebox{0.9}{$7.10 + 0.345\,\xi_* + 0.0832\,\xi_*^2$}&\scalebox{0.9}{$0.63 - 0.095\,\xi_* + 0.0048\,\xi_* ^2$}\\\hline
\cellcolor[gray]{0.9}\scalebox{0.9}{$\{+\mathcal{R}\mathcal{R}\}_{\rm M2}$} &\scalebox{0.9}{$-22.84 +15.17\, \xi_* + 0.0503\, \xi_* ^2$}&\scalebox{0.9}{$6.03 - 0.128\,\xi_* + 0.0748\, \xi_*^2$}&\scalebox{0.9}{$0.69 -0.100\,\xi_* + 0.0069\,\xi_* ^2 $}\\\hline
\hline
\end{tabular}
\caption{\label{tab:f3fit} The height $f^{c}_{3,j}$, location $x^c_{3,j}$ and width $\sgm_{3,j}$ of $f^{(3,\alpha)}_{h\mathcal{R}\mathcal{R}}$ \eqref{f3fit} for $\delta = 0.3$. For the large peak associated ($\alpha = L$) with the $j = -\mathcal{R}\mathcal{R}$ correlator and $j = +\mathcal{R}\mathcal{R}$, the fitting formulas are valid and $3.5\leq \xi_*\leq 6.5$ while for the small peak ($\alpha = S$) of $j = -\mathcal{R}\mathcal{R}$ correlator, they are valid for $4\leq \xi_*\leq 5$. The amplitude of $j = -\mathcal{R}\mathcal{R}$ correlator is positive $f^{c}_{3,j} > 0$ for the small peak and $f^{c}_{3,j} < 0$ for the large peak. The $j = +\mathcal{R}\mathcal{R}$ bispectrum has a single peak with a negative amplitude $f^{c}_{3,j} < 0$.}
\end{center}			
\end{table}

 A distinctive feature of the $ -\mathcal{R}\mathcal{R}$ correlator is its doubly peaked structure which occur with different signs and locations in $k$ space. In particular, $\langle \hat{h}_{-} \hat{\mathcal{R}} \hat{\mathcal{R}}\rangle$ exhibits a small positive peak $f^{c,S}_{3,j} > 0$ that occurs slightly earlier in $k$ space ($x^{c,S}_{3,j} < x^{c,L}_{3,j}$) compared to the following large peak realized in the opposite direction ($f^{c,S}_{3,j} < 0$). It is worth mentioning that such a feature is absent in the auto-correlators of sourced curvature $\mathcal{R}$ and ${h}$ metric perturbations \cite{Namba:2015gja,Ozsoy:2020ccy}. It would be interesting to investigate quantitatively whether the presence of such a small peak increase the observability of the TSS bispectrum. We present an analysis on the double peak structure of the $-\mathcal{R}\mathcal{R}$ correlator \eqref{f3MRR} by comparing it with $\mathcal{R} --$ (see below) correlator in Appendix \ref{AppC} where we show that the reason for this behavior stems from to the product of polarization vectors (see \eg eq. \eqref{pohv}) which serve the purpose of angular momentum conservation at each vertex in the diagrams of Figure \ref{fig:diag}. In particular, within the range of loop momenta where the gauge field sources have appreciable contribution to the $-\mathcal{R}\mathcal{R}$ diagram, we found that the product of polarization vectors in \eqref{pohv} have a sufficiently large both negative and positive peak depending on the orientation of the loop momentum (\ie non-planar vs. planar) with respect to the plane ($x$-$y$) where external momenta lives (See Figure \ref{fig:a}). Integrating over such configurations of the loop momentum (see eq. \eqref{f3MRR}) therefore yields to a double peaked structure that occur in opposite directions as we explain in detail in Appendix \ref{AppC1}. In what follows, in our discussion on the amplitude and shape of the TSS type bispectrum in Section \ref{S3p2}, we will focus our attention to the large peak that appears in Figure \ref{fig:f3fit} which constitutes the dominant scale dependent signal within the parameter space where $r_{s} \gg r_{\rm vac}$ (See Figure \ref{fig:rpeak}).

Another conclusion that can be drawn from Table \ref{tab:f3fit} and Figure \ref{fig:f3fit} is that Model 2 generically generates signals that has a smaller width compared to the Model 1 for the same parameter choice $\delta = 0.3$ ($\sgm^{(\rm M1)}_{3,j} > \sgm^{(\rm M2)}_{3,j}$) which in turn implies that the former requires a larger maximal value for the effective coupling $\xi_*$ between $\sgm$ and $A_\mu$ to generate a signal comparable in amplitude with Model 1. 

{\bf $\bullet\,\, +\mathcal{R}\mathcal{R}$ bispectrum:} On the other hand, we found that the $j =+\mathcal{R}\mathcal{R}$ correlator consist of a single peak that has the same Gaussian form as in \eqref{f3fit}. We provide the fitting formulas for this case in the third and sixth row in Table \ref{tab:f3fit}.  We see that due to the parity violation in the tensor sector, the  amplitude of $f^{(3)}_{+\mathcal{R}\mathcal{R}}$ is about an order of magnitude smaller than $j =-\mathcal{R}\mathcal{R}$ bispectrum. Note that this parity violation is not dramatic because $\langle \hat{h}_\lambda \hat{\mathcal{R}}\hat{\mathcal{R}}\rangle $ type non-Gaussianity contains only a single external state of the tensor perturbation with a definite polarization $\lambda  = \pm$. In general, we expect the parity violation in mixed 3-pt amplitudes to increase for an increasing number of external $h_\lambda$ in the bispectrum. In fact, as we will show, this is the case for the STT type bispectrum $\langle\hat{\mathcal{R}}\hat{h}_{\lambda}\hat{h}_\lambda\rangle$ below (See Section \ref{S3p1p2}).
\begin{table}
\begin{center}
\begin{tabular}{|c|c|c|c|}
\hline
\hline
\cellcolor[gray]{0.9}$\{j\}$&\cellcolor[gray]{0.9}$\ln(|f^c_{3,j}|)$&\cellcolor[gray]{0.9}$x^c_{3,j}$&\cellcolor[gray]{0.9}$\sgm_{3,j}$ \\
\hline
\cellcolor[gray]{0.9}\scalebox{0.95}{$\{\mathcal{R}--\}_{{\rm M}1}$}&\scalebox{0.9}{$-7.21 + 14.77\,\xi_*+ 0.117\,\xi_*^2$}&\scalebox{0.9}{$3.15 +0.665\, \xi_* + 0.0213\, \xi_* ^2$}&\scalebox{0.9}{$0.82 -0.109\, \xi_* + 0.0066\, \xi_* ^2$}\\\hline
\cellcolor[gray]{0.90}\scalebox{0.95}{$\{\mathcal{R}++\}_{{\rm M}1}$}&\scalebox{0.9}{$-15.6 + 14.74\,\xi_*+ 0.121\,\xi_*^2$}&\scalebox{0.9}{$1.24 +0.232\, \xi_* + 0.0147\, \xi_* ^2$}&\scalebox{0.9}{$0.80 -0.123\, \xi_* + 0.0085\, \xi_* ^2$}\\\hline\hline
\hline
\cellcolor[gray]{0.9}\scalebox{0.95}{$\{\mathcal{R}--\}_{{\rm M}2}$}&\scalebox{0.9}{$ -18.9 + 15.15\,\xi_*+ 0.0526\,\xi_*^2$} & \scalebox{0.9}{$6.60 +0.050\, \xi_* + 0.0731\, \xi_* ^2$}&\scalebox{0.9}{$\,\,0.65 - 0.089\,\xi_* + 0.0058\, \xi_* ^2$ }\\\hline
\cellcolor[gray]{0.9}\scalebox{0.95}{$\{\mathcal{R}++\}_{{\rm M}2}$}&\scalebox{0.9}{$ -27.7 + 15.25\,\xi_*+ 0.0484\,\xi_*^2$} & \scalebox{0.9}{$1.92 +0.290\, \xi_* + 0.0083\, \xi_* ^2$}&\scalebox{0.9}{$\,\,0.57 - 0.064\,\xi_* + 0.0034\, \xi_* ^2$ }\\\hline
\hline
\end{tabular}
\caption{\label{tab:f3fit2} Fitting formulas for the height $f^{c}_{3,j}$, location $x^c_{3,j}$ and width $\sgm_{3,j}$ that parametrize the scale dependent enhancement of the $j = \{\mathcal{R}\lambda\lambda\}$ type mixed bispectrum in \eqref{SC3pt}. Formulas are obtained for $\delta = 0.3$ and $3.5\leq \xi_*\leq 6.5$.}
\end{center}			
\end{table}\noindent
\begin{figure}[t!]
\begin{center}
\includegraphics[scale=0.87]{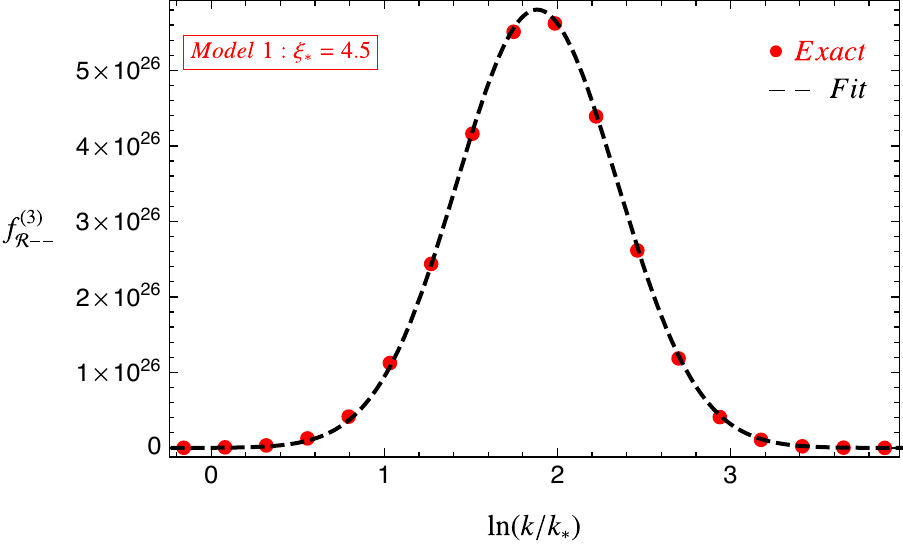}\includegraphics[scale=0.87]{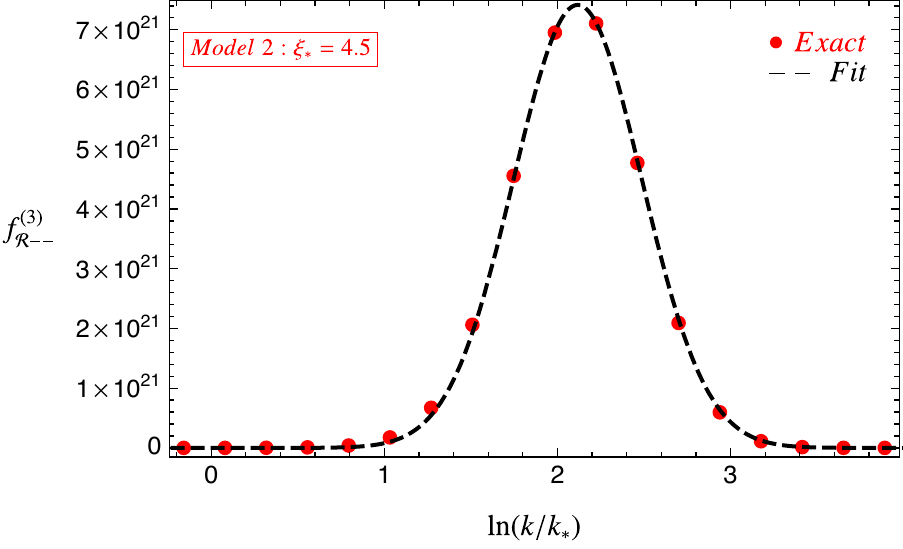}
\end{center}
\caption{The scale dependence of $f^{(3)}_{\mathcal{R}--}$ for Model 1 (Left) and Model 2 (Right) described by the potentials \eqref{pots} in the spectator axion-gauge field model \eqref{Lm}. The red dots are obtained by numerical evaluation of \eqref{f3RMM} at $x_2 = x_3$ for a grid of $x_* = k/k_*$  values and dashed lines show the accuracy of the Gaussian expression \eqref{f3fit} where we utilized Table \ref{tab:f3fit2}. \label{fig:f3fit2}}
\end{figure}
\subsubsection{STT correlator}\label{S3p1p2}
Repeating the analysis we performed for the TSS type correlator, we found that the scale dependence of $f^{(3)}_{\mathcal{R}\lambda\lambda}$ in \eqref{f3RMM} can instead be described by a single Gaussian peak (See Appendix \ref{AppC1}) that takes the same form as in \eqref{f3fit}. For $3.5 \leq \xi_* \leq 6.5$, its height and width and location can be well fitted by the second order formulas we provide in Table \ref{tab:f3fit2} and the accuracy of these formulas compared to the exact numerical computation of \eqref{f3RMM} is shown in Figure \ref{fig:f3fit2}. From Table \ref{tab:f3fit2}, we see that parity violation present itself stronger for the STT compared to the TSS bispectrum as expected since it has more external tensor mode that carry a definite polarization. On the other hand, $\mathcal{R}h_\lambda h_\lambda$ carries similar features with TSS correlator, such as the width of the signal in Model 1 is larger than the second which in turn imply that Model 2 requires a larger effective coupling $\xi_*$ to generate the same amount of signal. This situation appears to hold generically for any correlators containing observable fluctuations $\mathcal{X} =\{\mathcal{R},h_{\lambda}\}$ and stems from the fact that in the second Model (M2), spectator axion probes a sharper region of its potential (\ie cliff like regions) compared to Model 1, leading to the excitation of a smaller number of gauge field modes when the particle production is maximal, \ie around $\tau \sim \tau_*$.
\subsection{Amplitude and shape dependence of mixed non-Gaussianities}\label{S3p2}
Having studied the scale dependent amplification of non-Gaussian signals of mixed type, in this section we investigate their amplitude and shape. 

\smallskip
\noindent
{\bf Amplitude of the bispectra.} To quantify the size of the mixed non-Gaussianity, we will make use of the standard definition the non-linearity parameter evaluated at the equilateral configuration \cite{Komatsu:2001rj,Barnaby:2011vw}, 
\beq\label{fnl}
f^{\,j}_{\rm NL} (k) = \fr{10}{9}\fr{k^6}{(2\pi)^{5/2}} \fr{\mathcal{B}^{(s)}_{\,\,\,j} (\vec{k},\vec{k},\vec{k})}{\mathcal{P}^2_{\mathcal{R}}(k)},
\eeq
\noindent
where we restrict our analysis with the dominant correlators, $j = \{\mathcal{R}--, -\mathcal{R}\mathcal{R}\}$ (See Table \ref{tab:f3fit} and \ref{tab:f3fit2}). As we showed in the previous section, the transient particle production in the gauge field sector leads to a scale dependent bump in the 3-pt correlators of mixed type. To estimate the maximal size of the non-linearity parameters $f^{\,j}_{\rm NL}$, we therefore use \eqref{SC3pt} to evaluate \eqref{fnl} at the peak of the sourced GW signal, $k = k_* x^{c}_{2,-}$ (See Table \ref{tab:fit1}). 

\begin{figure}[t!]
\begin{center}
\includegraphics[scale=0.87]{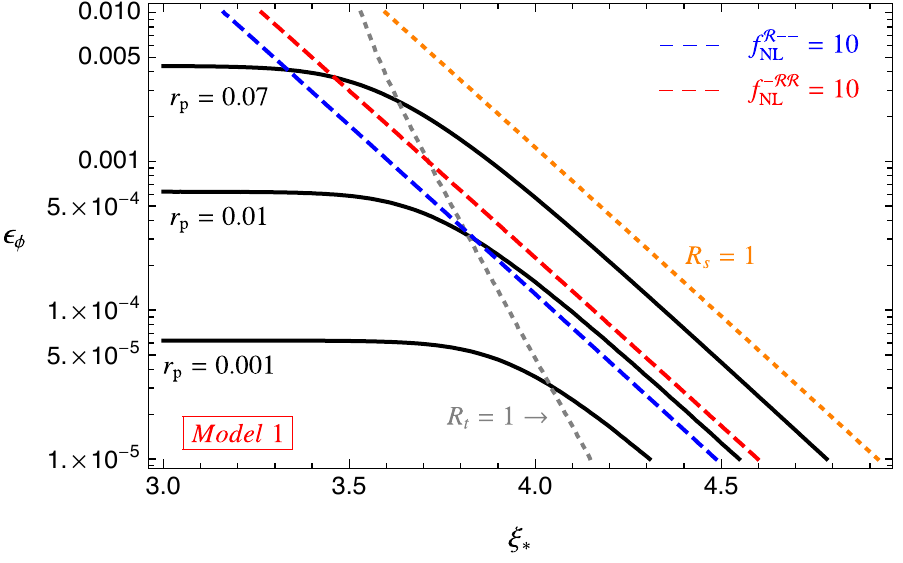}\includegraphics[scale=0.87]{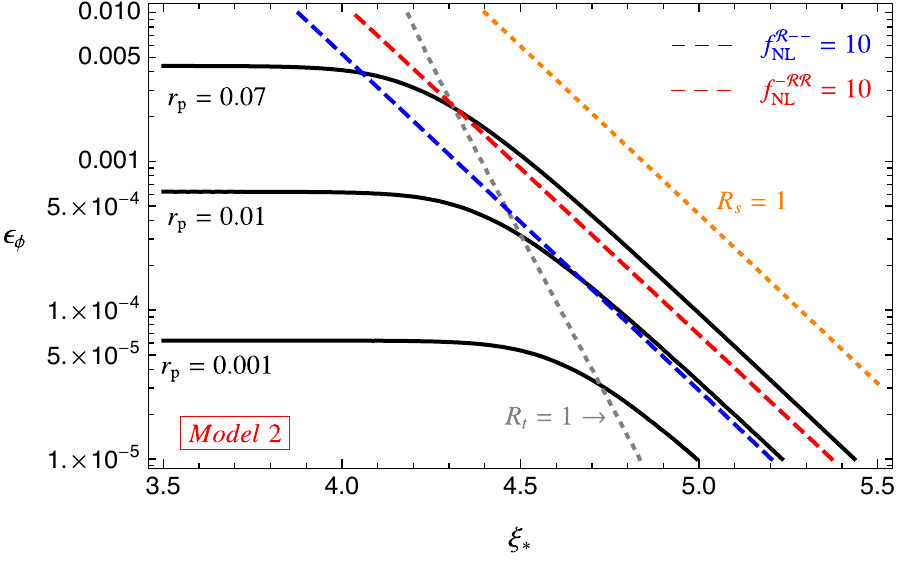}
\end{center}
\caption{Constant $f^{j}_{\rm NL}$ curves superimposed with constant $r$ curves in the $\epsilon_\phi -\xi_*$ plane for Model 1 (Left) and Model 2 (Right). The color coding and the parameter choices are the same as in Figure \ref{fig:rpeak}.\label{fig:rvsfnl}}
\end{figure}

To visualize the relevant parameter space where mixed non-Gaussianity is significant, in Figure \ref{fig:rvsfnl}, we plot $f^j_{\rm NL} =10$ curves in the model parameter space ($\epsilon_\phi -\xi_*$). We see that the parameter space where GW's sourced by the gauge field sources dominate (on the right hand side of $R_t = 1$ line in Figure \ref{fig:rvsfnl}) overlaps with the sizeable values of $f^{j}_{\rm NL}$. In this regime, $f^{j}_{\rm NL}$ can be parametrized in terms of the peak value of the tensor-to-scalar ratio $r_{\rm p}$ \eqref{rpeak} as
\noindent  
\beq\label{fnlpeakrhh}
f^{\mathcal{R}--}_{\rm NL} \simeq 
 \begin{dcases} 
       24 \left(\fr{r_{\rm p}}{0.01}\right)^{3/2}\, e^{0.025\xi_*}\quad({\rm Model}\,\,1)\\
       11  \left(\fr{r_{\rm p}}{0.01}\right)^{3/2}\, e^{0.082\xi_*}\quad({\rm Model}\,\, 2),
   \end{dcases}
\eeq 
and
\beq\label{fnlpeakhrr}
f^{-\mathcal{R}\mathcal{R}}_{\rm NL} \simeq 
 \begin{dcases} 
       8\,\, \left(\fr{r_{\rm p}}{0.01}\right)^{3/2}\,\,\, e^{-0.099\xi_*}\quad({\rm Model}\,\,1)\\
       1.6  \left(\fr{r_{\rm p}}{0.01}\right)^{3/2}\, e^{-0.009\xi_*}\quad({\rm Model}\,\, 2).
   \end{dcases}
\eeq 
where $\xi_*$ dependence is weak and hence can be ignored for the parameter space of interest $3.5 \leq \xi_* \leq 6.5$. 
The origin of $f^j_{\rm NL} \propto r_{\rm p}^{3/2}$ scaling for STT and TSS correlators can be understood as follows. In the effective particle production regime we are interested in ($ R_t > 1$), tensor power spectrum arise through a loop diagram that contains two copies of three legged vertex in Figure \ref{fig:diag}, \ie $(h A A)^2 \propto A^6$ and thus carries a weight factor of $(e^{c \pi \xi_*})^4$ (See Table \ref{tab:fit0}) that characterize the amplification of gauge modes by the transiently rolling spectator axion. Therefore, evaluated at the peak of scale dependent signals, one loop diagram involving two external gravitons gives $\mathcal{P}_h \propto r_{\rm p} \propto e^{4 c \pi \xi_*}$ for $R_s < 1$. On the other hand, we notice from Figure \ref{fig:diag} that diagrams that contribute to TSS and STT correlators originate from the fusion of three point tensor and scalar vertices of the following form $(hAA)^2 \times ([\delta \phi\leftarrow\delta\sgm] AA)$ and $(hAA) \times ([\delta \phi\leftarrow\delta\sgm] AA)^2$ and thus we roughly have $f^j_{\rm NL} \propto \mathcal{B}^j \propto A^6 \propto e^{6c\pi \xi_*}$ for $R_s < 1$. Putting together the arguments above thus gives the scaling $f^j_{\rm NL} \propto r_{\rm p}^{3/2}$. 

At this point, it is useful to compare these results with the TSS and STT type non-Gaussianity obtained in single field inflation where $f^{{\mathcal{R}--}}_{\rm NL} \sim r^{2}$ and $f^{-\mathcal{R}\mathcal{R}}_{\rm NL} \sim r$ is expected \cite{Maldacena:2002vr}. In this respect, the results in \eqref{fnlpeakrhh} and \eqref{fnlpeakhrr} can be considered as a new set of consistency conditions that can be utilized the distinguish particle production scenarios involving Abelian gauge fields from the conventional ones. In particular, these results indicate that the gauge field production induced by the rolling axions can clearly alter the parametric dependence of non-linearity parameters on $r$ and STT and TSS mixed non-Gaussianity shows a significant enhancement with respect to the standard results from single field inflation. 

Furthermore, the relative locations of $f^{-\mathcal{R}\mathcal{R}}_{\rm NL}$ and $f^{\mathcal{R}--}_{\rm NL}$ curves in Figure \ref{fig:rvsfnl} indicated that the non-Gaussianity associated with the latter is larger for a given $\xi_*$ that parametrizes the strength of gauge field production. On the other hand, a comparison between Table \ref{tab:f3fit} and \ref{tab:f3fit2} reveals that the level of parity violation is more emphasized for the STT bispectrum compared to TSS. As we mentioned above, this result is expected since STT type correlator contains more external tensor mode with a definite parity $\lambda = \pm$.

\begin{figure}[t!]
\begin{center}
\includegraphics[scale=0.82]{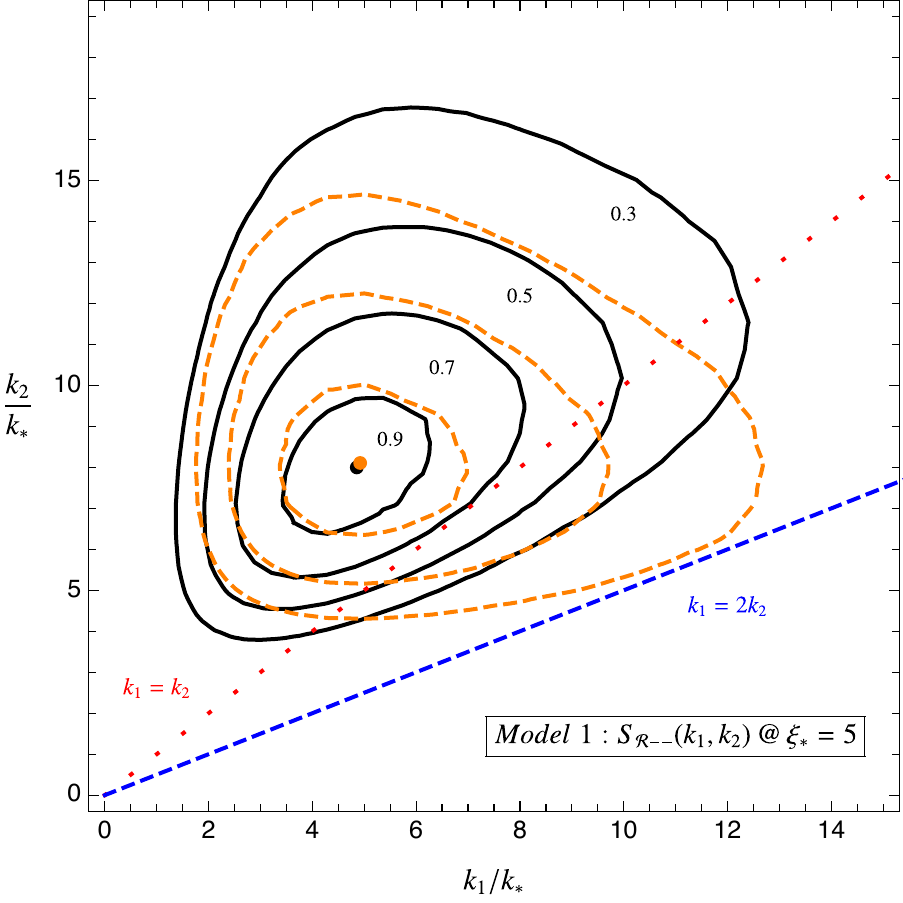}\,\,\,\,\includegraphics[scale=0.82]{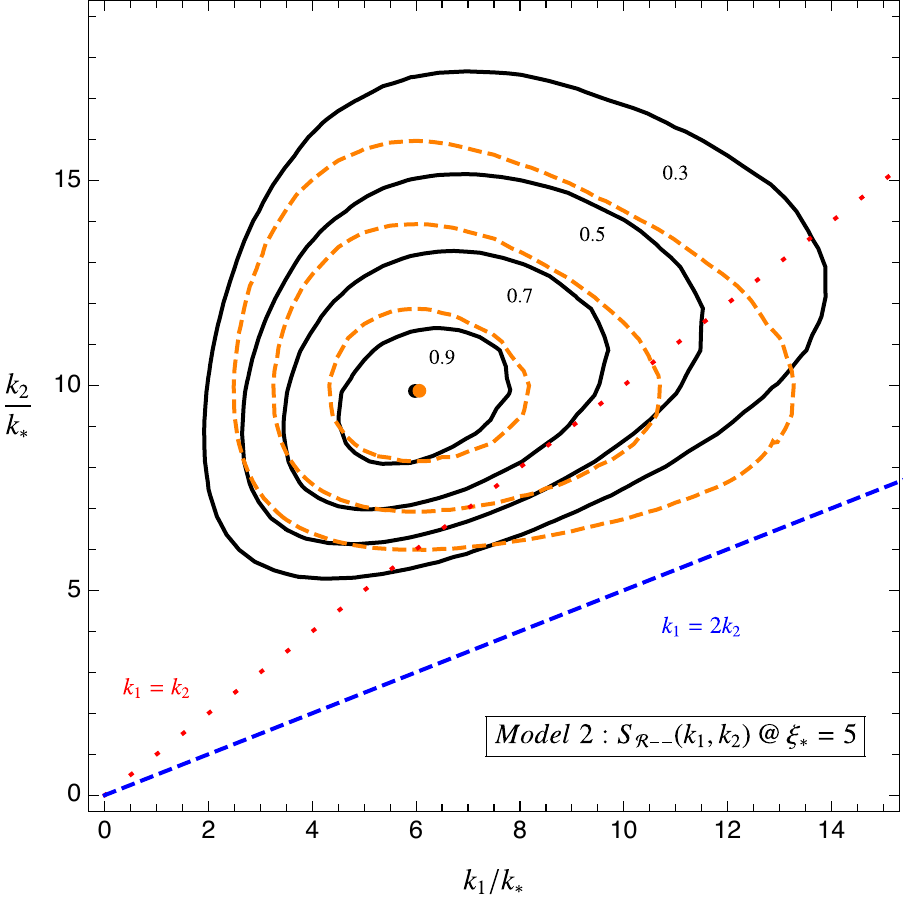}\\\smallskip
\includegraphics[scale=0.82]{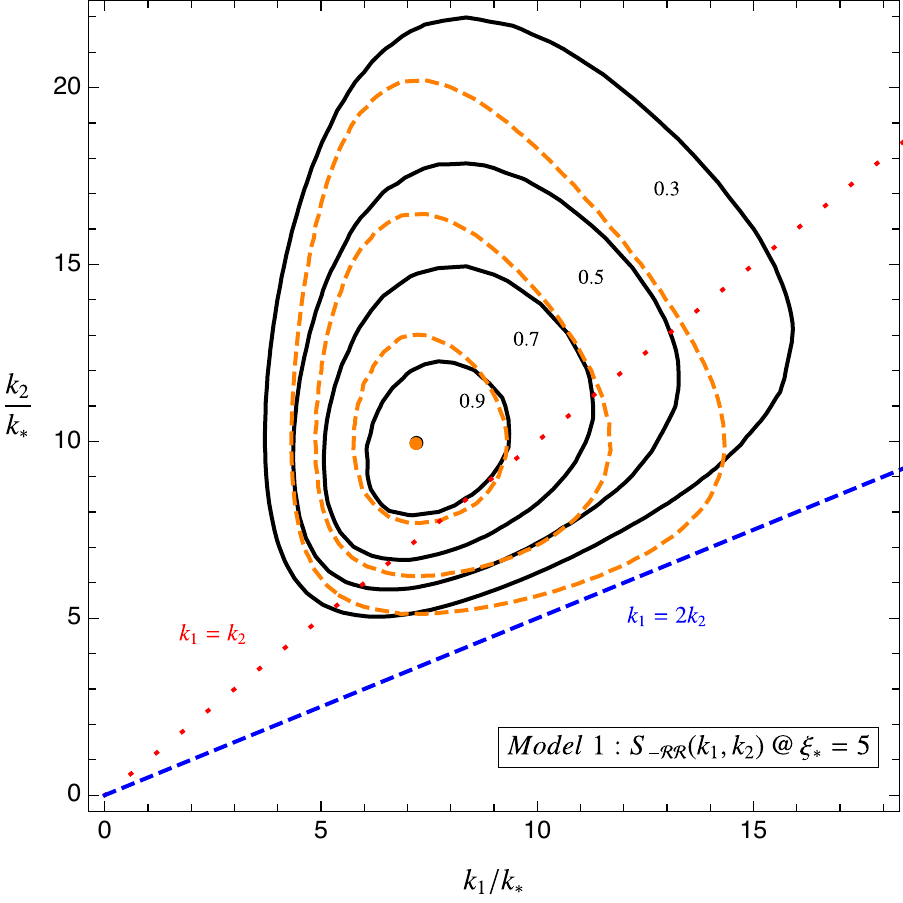}\,\,\,\,\includegraphics[scale=0.82]{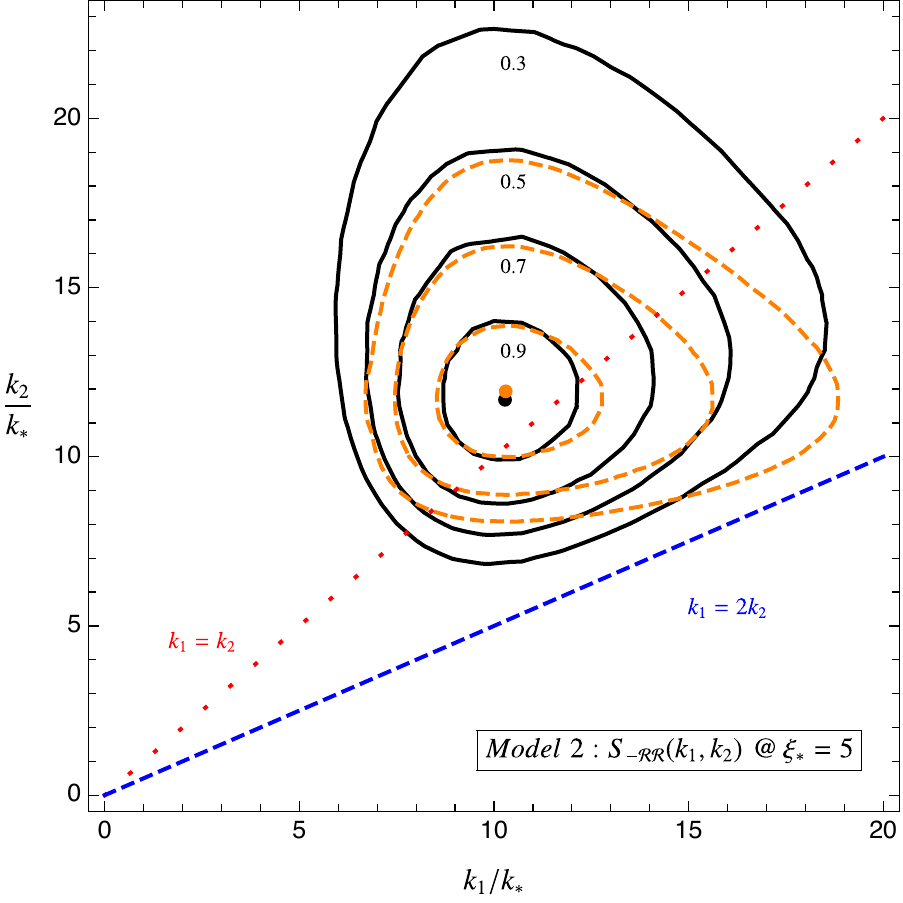}
\end{center}
\caption{Constant contour lines of the shape $S_{\mathcal{R}--}$ (top panel) and $S_{-\mathcal{R}\mathcal{R}}$ (bottom panel) in the $k_1/k_* - k_2/k_*$ plane, respectively for the TSS and STT mixed bispectrum and for both rolling axion models we consider in this work. The black dots locate the triangle configuration for which bispectrum is maximum ($S_{j} = 1$). Orange dashed lines indicate the $S_j = 0.9,0.7,0.5$ contour lines derived from the approximate expressions \eqref{f3rmmapp} (top) and \eqref{f3mrrapp} (bottom) while the orange dot locates the triangle configuration of their corresponding maximum. The red-dotted line and the blue dashed line refers to equilateral triangles $k_1 = k_2$ and folded triangles $k_1 = 2 k_2$ respectively. \label{fig:S}}
\end{figure}

\smallskip
\noindent
{\bf Shape of the mixed bispectra.} We now turn our attention to the shape of the mixed non-Gaussianity. For this purpose, it is customary to extract the overall $k^{-6}$ scaling of the bispectrum in \eqref{SC3pt} by defining the shape function $S_j$ as \cite{Babich:2004gb,Fergusson:2008ra},
\beq\label{sf}
S_{j} (k_1,k_2,k_3) = \mathcal{N} (k_1 k_2 k_3)^2 \mathcal{B}^{(\rm s)}_{j} (\vec{k}_1,\vec{k}_2,\vec{k}_3)\propto  f^{(3)}_{\,\,\,j}\left(\xi_{*}, x_{*}, \delta, x_{2}, x_{3}\right)
\eeq
where $\mathcal{N}$ is an arbitrary normalization factor. We pick a normalization factor $\mathcal{N}$ to ensure $S_j = 1$ at the triangle configuration $f^{(3)}_j$ becomes maximal (obtained numerically). Then focusing on isoceles triangles $k_2 = k_3$ ($x_2 = x_3$), we evaluate the shape function on a grid of values in the $k_1/k_* - k_2/k_*$ plane for $j =\{\mathcal{R}--, -\mathcal{R}\mathcal{R}\}$ and plot the resulting constant contour lines corresponding to $S_j = \{0.9, 0.7, 0.5, 0.3\}$ values of the shape function in Figure \ref{fig:S}. We note that, due to the triangle inequality $k_2 + k_3 \geq k_1$, only triangle configurations that satisfy $k_1 \leq 2k_2$ ($x_2 \geq 1/2$) is allowed as shown by the limiting blue dashed lines shown in Figure  \ref{fig:S}.

We observe from the shape of the contour lines (such as their spread in the $k_1-k_2$ plane and the approximate locations of the maximum) that both of the rolling axion models lead to qualitatively similar result for each type of mixed bispectrum. A slightly different behavior appears for TSS correlator in the rolling axion monodromy model (M2) where the spread of the contour lines take up a smaller area in the $k_1 - k_2$ plane. The reason for this is the fact the for the same parameter choices the second model contains a sharper feature in its dynamics compared to the M1. In particular, the physical quantity that controls the gauge field production, \ie the velocity of the $\dot{\sgm}$ have a more spiky behavior in the second Model, leading to the excitation of a smaller range of gauge field modes that can in turn source tensor and scalar fluctuations. This effect becomes more emphasized as the number of external $\mathcal{R}$ in the 3-pt function increases because the sourced curvature perturbation is more susceptible to the background evolution of the rolling axion as can be verified explicitly by comparing eqs. \eqref{CPSapp}-\eqref{ts}. The similarity of the shapes of the contour lines in the STT correlators (top row) in Figure \ref{fig:S} also supports these arguments. 

As indicated by the location of the black dots in Figure \ref{fig:S}, a common feature of the mixed correlators is that they are maximized for triangle configurations close to the equilateral shape. In particular, away from the maximum, the shape function reduces considerably in magnitude towards the folded $k_1/k_* \to 2 k_2/k_*$ and squeezed configurations $k_1/k_* \to 0$, implying that the shape of the STT and TSS bispectra are distinct from such configurations. Besides the general features we covered so far, there are some quantitative differences in the properties of the shape functions which we discuss in detail below. 

{\bf $\bullet$ STT:} From the top panel in Figure \ref{fig:S}, notice that mixed correlators are maximal at scales that slightly deviates from the exact equilateral configuration, \ie $S_{\mathcal{R}--} = 1$ (black dots) at $k_2 > k_1$ for both models.

This slight deviation from the exact equilateral configuration is closely tied to the offset that appear in the locations of the sourced curvature perturbation $\mathcal{R}$ and $h_-$. In particular, a close inspection of the peak location of the 2-pt correlators (See Table \ref{tab:fit1}) reveals $x^c_{2,\mathcal{R}} < x^c_{2,-}$ and naturally leads to the expectation that the peak location of the external momenta ($k_1$) associated with $\mathcal{R}$ to satisfy $k_1 < k_2$ in the STT correlator.

{\bf $\bullet$ TSS:} From the bottom panels of Figure \ref{fig:S}, we notice that the TSS bispectrum also takes its maximal value for triangles satisfying  $k_2 > k_1$.Considering the discussion we presented above for the STT correlator above, this result contradicts with the expectation that the maximum of the TSS bispectra should appear below the equilateral line $k_2 < k_1$ in Figure \ref{fig:S}. We speculate that this peculiarity stems from the products of helicity vectors $\epsilon_{(-\mathcal{R}\mathcal{R})}$ (defined in \eqref{pohv}) that appear inside the integral \eqref{f3MRR} of $f^{(3)}_{-\mathcal{R}\mathcal{R}}$ that characterize the shape function $S_{-\mathcal{R}\mathcal{R}}\propto f^{(3)}_{-\mathcal{R}\mathcal{R}}$, which should have more support for $x_2 > 1$ ($k_2 > k_1$).

To speed up its analysis with the actual data, in Appendix \ref{AppD}, we derive approximate expressions for the mixed bispectra, written as a sum of factorized terms. In particular, we provide approximate expressions for the bispectra such that each term in these expressions (see \eg \eqref{f3rmmapp} and \eqref{f3mrrapp}) is written as a product of source functions $f_{2,j}$ and $f^{(3)}_j$ that depend on only one external momenta, $k_i$. In Figure \ref{fig:S}, we illustrate the accuracy of these approximate expressions by the orange dashed lines and orange dots corresponding to $S_j = 0.9,0.7,0.5$ contour lines and $S_j =1$ point, respectively. As shown in the Figure, we observe that the maximums derived from the approximate formulas (orange dots) are nearly coincident with the actual ones (black dots) and the approximate expressions provide an accurate description of the exact bispectra, particularly around the maximum. 

\section{Conclusions}\label{S4}
In this work, we focused our attention to a class of inflationary scenarios characterized by a system \eqref{Lm} of spectator axion and $\rm U(1)$ gauge fields coupled by a Chern-Simons type interaction \cite{Namba:2015gja,Ozsoy:2020ccy}. In these models, the transient roll of spectator axion triggers a localized enhancement in the gauge field fluctuations which in turn induces several phenomenologically interesting signatures in the CMB observables, including a low energy scale realization of inflation endowed with scale dependent chiral GW signal accessible by forthcoming observations \cite{Namba:2015gja} together with observable tensor non-Gaussianity \cite{Shiraishi:2016yun}. While bounds on the scalar 2-pt and 3-pt auto-correlators from the CMB observables can be avoided for the parameter space that leads to interesting phenomenology in the tensor sector, the spectator axion- $\rm U(1)$ gauge field dynamics also predict enhanced scalar fluctuations by the gauge fields. Therefore in this setup, mixed non-Gaussianities including scalar and tensor fluctuations appear to be as important as the information one can gain from their auto-correlators. In particular, the non-trivial parity violating structure of these correlators may provide additional predictive power, help us constrain the model parameters and reveal distinguishing features that can provide effective model comparison, \eg considering the absence/presence of the analogue signals that are present within the standard single field inflation or the close $\rm SU(2)$ cousin \cite{Dimastrogiovanni:2016fuu,Fujita:2017jwq} of the model we consider in this work. 

To shed some light on these issues, we derived predictions for the scalar-tensor-tensor and tensor-scalar-scalar bispectrum focusing on spectator axion-$\rm U(1)$ gauge field dynamics during inflation (See Section \ref{S3}). We find that both bispectra exhibit a scale dependent amplification and at the their respective peaks they are significantly enhanced compared to their counterparts in the minimal single field inflationary scenario (See \eg eqs. \eqref{fnlpeakrhh} and \eqref{fnlpeakhrr}). In particular, in the efficient particle production regime, we found that mixed non-Gaussian correlators satisfy a new consistency condition: $f^j_{\rm NL} \simeq \mathcal{O}(1-10) (r/0.01)^{3/2}$ that distinguishes these models from the conventional single field scenarios. More importantly, due to parity violating nature of gauge field production, the resulting mixed bispectra also exhibit a preferred chirality. 
In section \ref{S3p2}, we studied the shape dependence of these amplified signals and found that both bispectra is maximal \emph{close} to the equilateral shape, slightly deviating from the exact equilateral configuration (See Figure \ref{fig:S}). Given that the sources (gauge fields) of these correlators have maximal amplitudes at around horizon crossing (See Table \ref{tab:fit1}), this result is expected.  
   
The detectability of the scale dependent, parity violating $\langle\hat{\mathcal{R}}\hat{h}_\lambda\hat{h}_\lambda\rangle$ and $\langle\hat{h}_\lambda\hat{\mathcal{R}}\hat{\mathcal{R}}\rangle$ signals we studied in this work require a detailed analysis in the CMB observables. A possibility in this direction would be to consider cross correlations between the CMB temperature {\rm T} and {\rm E,B} polarization modes. In particular, to search for a primordial STT bispectrum, a suitable observable would be TBB and EBB cross correlations of the CMB (See \eg \cite{Shiraishi:2012sn,Shiraishi:2013vha,Bartolo:2018elp}) whereas the observability of TSS bispectrum can be analyzed through BTT or BEE (See \eg \cite{Domenech:2017kno}). In this respect, the approximate factorized expressions we derived for the both bispectra (See Appendix \ref{AppD}) can be utilized to test mixed non-Gaussian signals of the rolling spectator axion models. On the other hand, although we focus much of our attention on the impact of scalar and tensor cross correlations at CMB scales in this work, rolling spectator axion models can also produce interesting signals at much smaller cosmological scales (See \eg \cite{Ozsoy:2020ccy,Garcia-Bellido:2016dkw}). In this context, it would be interesting to explore observables that parity violating STT and TSS correlators may induce at sub-CMB scales. We leave further investigations on these issues for a future publication. 
\acknowledgments
We would like to thank Marco Peloso, Maresuke Shiraishi and Caner \"Unal for comments and useful discussions pertaining this work. Part of this research project was conducted using computational resources at Physics Institute of the Czech Academy of Sciences (CAS) and we acknowledge the help of Josef Dvoracek on this process. This work is supported by the European Structural and Investment Funds and the Czech Ministry of Education, Youth and Sports (Project CoGraDS-CZ.02.1.01/0.0/0.0/15003/0000437).
\begin{appendix}
\section{Gauge field modes as sources of scalar and tensor perturbations}\label{AppA}
We now summarize some important aspects of the gauge field production and their subsequent sourcing of cosmological perturbations. 
Considering the time dependent profile  \eqref{Joep} of the effective coupling $\xi_*$, Eq. \eqref{nmea} describes the standard Schr\"odinger equation of the ``wave-function" $A_-$ for which an analytic solution can be derived by employing WKB approximation methods \cite{Namba:2015gja}. In particular, the late time growing solution to the eq. \eqref{nmea} can be parametrized in terms of a scale dependent normalization (real and positive) factor as \cite{Namba:2015gja,Ozsoy:2020ccy}:
\beq\label{solapp}
A_{-}(\tau, k) \simeq \fr{N_A(\xi_*, -k\tau_*, \delta)}{\sqrt{2k}} \left[\frac{-k\tau}{2 \xi(\tau)}\right]^{1 / 4}  \exp \left[- E(\tau){\sqrt{-2\xi_*k\tau} } \right]\quad\quad\quad  \tau/\tau_* < 1,
\eeq
where the time dependent argument of the exponential factor depends on the model as 
\beq\label{Q}
E(\tau)=
 \begin{dcases} 
       \frac{2\sqrt{2}}{(1+\delta) (\tau/\tau_*)^{-\delta/2}},& \quad{\rm Model\, 1}\,({\rm M}1) \,,\\
       \fr{2}{\delta\, |\ln(\tau/\tau_*)\,|},&\quad{\rm Model\, 2}\,({\rm M}2).
   \end{dcases}
\eeq 
The scale dependence ($x_* = k/k_*$) of the normalization factor $N_A(\xi_*, x_*, \delta)$ in \eqref{solapp} can be determined by solving \eqref{nmea} numerically for different values of $x_* = -k\tau_*$ and matching it to the WKB solution \eqref{solapp} at late times $-k\tau \ll 1$. In this way, one can confirm that that $N_A(\xi_*, x_*,\delta )$ can be accurately described by a log-normal distribution,
\beq\label{Nform}
N_A\left(\xi_{*}, x_*, \delta\right) \simeq N_A^{c}\left[\xi_{*}, \delta\right] \exp \left(-\frac{1}{2 \sigma_A^{2}\left[\xi_{*}, \delta\right]} \ln ^{2}\left(\frac{x_*}{q_A^{c}\left[\xi_{*}, \delta\right]}\right)\right),
\eeq
where the functions $N_A^{c}, q_A^c$ and $\sgm_A$ parametrizes the background dependence of gauge field production, and hence depend on $\xi_*$ and $\delta$. For an effective coupling to gauge fields within the range $3 \leq \xi_* \leq 6.5$, these functions can be described accurately by a second order polynomial in $\xi_*$ provided in Table \ref{tab:fit0}.
\begin{table}
\begin{center}
\begin{tabular}{|c |c |c |c |}
\hline
\hline
\cellcolor[gray]{0.9}&\cellcolor[gray]{0.9}$\ln(N_A^{c})$&\cellcolor[gray]{0.9}$q_A^{c}$&\cellcolor[gray]{0.9}$\sgm_A$ \\
\hline
\cellcolor[gray]{0.9}\scalebox{0.95}{${{\rm M}1}$}&\scalebox{0.9}{$0.290 + 2.83\,\xi_*+ 0.00100\,\xi_*^2$}&\scalebox{0.9}{$-0.097 +0.633\, \xi_* - 0.00110\, \xi_* ^2$}&\scalebox{0.9}{$2.11 -0.321\, \xi_* + 0.0208\, \xi_* ^2$}\\\hline
\cellcolor[gray]{0.9}\scalebox{0.95}{${{\rm M}2}$}&\scalebox{0.9}{$ 0.325+2.72 \,\xi_{*}-0.00069 \,\xi_{*}^{2}$} & \scalebox{0.9}{$0.013 + 0.710 \,\xi_{*}-0.00105\, \xi_{*}^{2}$}&\scalebox{0.9}{$1.69-0.254\, \xi_{*} + 0.0164\, \xi_{*}^{2}$} \\\hline
\hline
\end{tabular}
\caption{\label{tab:fit0} $\xi_*$ dependence of the functions $N_A^{c}$, location $q_A^c$ and width $\sgm_A$ that appear in the late time amplitude $N_A$ (eq. \eqref{Nform}) of the gauge field modes for $\delta = 0.3$ and $3\leq \xi_*\leq 6.5$.}
\end{center}			
\end{table}
Since the growing solution to the mode functions $A_-$ is real, its Fourier decomposition can be simplified as,
\beq\label{DGFs}
\hat{A}_i(\tau, \vec{x}) \simeq \int \fr{\d^3 k}{(2\pi)^{3/2}} ~ e^{i\vec{k}.\vec{x}}  \epsilon^{-}_i(\vec{k}) A_{-}(\tau,\vec{k})\left[
\hat{a}_{-}(\vec{k}) + \hat{a}^{\dagger}_{-}(-\vec{k})  \right], 
\eeq
where the helicity vectors obey $k_i \epsilon^{\pm}_i = 0$, $\epsilon_{ijk}~ k_j ~\epsilon^{\pm}_k = \mp i k \epsilon^{\pm}_i$, $\epsilon^{\pm}_i\epsilon^{\pm}_i = 0$, $\epsilon^{\pm}_i \epsilon^{\mp}_i =1$ and $(\epsilon^{\lambda}_i(\vec{k}))^{*} = \epsilon^{\lambda}_i(-\vec{k}) = \epsilon^{-\lambda}_i(\vec{k})$ and the annihilation/creation operators satisfy $\left[\hat{a}_\lambda(\vec{k}),\hat{a}^\dagger_{\lambda'}(\vec{k}')\right] = \delta_{\lambda\lambda'} \,\, \delta(\vec{k}-\vec{k}')$. Using the definitions of ``Electric'' and ``Magnetic" from the main text (See below eq. \eqref{te}), we obtain their Fourier modes as
\begin{align}\label{EBF}
\nn \hat{E}_i(\tau,\vec{k}) &= - \sqrt{\fr{k}{2}}\,\fr{\epsilon_{i}^{-}(\vec{k})}{a(\tau)^{2}} \left(\fr{2\xi(\tau)}{-k\tau}\right)^{1/4}  N_A(\xi_*,-k\tau_*,\delta) \exp \left[-E(\tau)\sqrt{-2\xi_*k\tau}  \right]\hat{\mathcal{O}}_{-}(\vec{k}),\\
\hat{B}_i(\tau,\vec{k}) &= -\sqrt{\fr{k}{2}}\,\fr{\epsilon_{i}^{-}(\vec{k})}{a(\tau)^{2}}\left(\fr{-k \tau}{2\xi(\tau)}\right)^{1/4} N_A(\xi_*,-k\tau_*,\delta) \exp \left[-E(\tau){\sqrt{-2\xi_*k\tau}}\right]\hat{\mathcal{O}}_{-}(\vec{k}),
\end{align}
where we defined the following shorthand notation for the superposition of gauge field annihilation and creation operators: $\hat{\mathcal{O}}_\lambda(\vec{q}) \equiv \left[\hat{a}_{\lambda}(\vec{q})+\hat{a}_{\lambda}^{\dagger}(-\vec{q})\right]$. 

Electric and Magnetic fields defined in \eqref{EBF} act as sources to cosmological scalar and tensor perturbations. The main channel of contribution to curvature perturbation in this model is schematically given by  $\delta A + \delta A \to \delta \sgm \to \delta \phi \propto \mathcal{R}$  \cite{Ferreira:2014zia,Ozsoy:2017blg} and can be expressed as \cite{Namba:2015gja,Ozsoy:2020ccy}
\beq\label{CPSapp}
\hat{\mathcal{R}}^{(s)}(\tau, \vec{k}) \simeq \frac{3 \sqrt{2} H \tau}{\Mp} \int d \tau^{\prime} G_{k}\left(\tau, \tau^{\prime}\right) \frac{\sqrt{\epsilon_{\sigma}\left(\tau^{\prime}\right)}}{\tau^{\prime 2}} \int d \tau^{\prime \prime} G_{k}\left(\tau^{\prime}, \tau^{\prime \prime}\right) \hat{J}_{\sigma}\left(\tau^{\prime \prime}, \vec{k}\right),
\eeq
where $G_k$ is the Green's function for the operator $\partial_\tau^2 + k^2 -2/\tau^2$ and the source term $\hat{J}_\sgm$ is given by
\beq\label{Jps}
\hat{J}_\sgm(\tau'',\vec{k}) =\fr{\alpha_{\rm c}a(\tau'')^3}{f}\int\fr{\d^3 p}{(2\pi)^{3/2}}~ \hat{E}_{i}(\tau'', \vec{k}-\vec{p}) ~\hat{B}_{i}(\tau'',\vec{p}).
\eeq

On the other hand, metric fluctuations are inevitably sourced by the traceless transverse part of the anisotropic energy momentum tensor as
\beq\label{sh}
\hat{h}^{(s)}_\lambda (\tau, k) = \fr{2}{a(\tau)\Mp} \int_{-\infty}^{\tau} \d\tau'~ G_k(\tau,\tau')~ \hat{J}_\lambda(\tau',\vec{k}),
\eeq
where $J_\lambda$ can be expressed as a bilinear convolution of electric and magnetic fields:
\beq\label{ts}
\hat{J}_{\lambda}(\tau, \vec{k})=-\frac{a^{3}(\tau)}{M_{\mathrm{pl}}} \Pi_{i j, \lambda}(\vec{k}) \int \frac{\mathrm{d}^{3} p}{(2 \pi)^{3 / 2}}\left[\hat{E}_{i}(\tau, \vec{k}-\vec{p}) \hat{E}_{j}(\tau,\vec{p})+\hat{B}_{i}(\tau,\vec{k}-\vec{p}) \hat{B}_{j}(\tau,\vec{p})\right],
\eeq
with  $\Pi_{ij,\pm} = \epsilon^{\mp}_i(\vec{k}) \epsilon^{\mp}_j(\vec{k})$ is the transverse traceless projector with the properties listed below \eqref{te}.
\section{Fitting formulas for $f_{2,j}$ and the tensor to scalar ratio}\label{AppB}
\begin{table}
\begin{center}
\begin{tabular}{|c |c |c |c |}
\hline
\hline
\cellcolor[gray]{0.9}$\{i,j\}_{\alpha}$&\cellcolor[gray]{0.9}$\ln(f^c_{i,j})$&\cellcolor[gray]{0.9}$x^c_{i,j}$&\cellcolor[gray]{0.9}$\sgm_{i,j}$ \\
\hline
\cellcolor[gray]{0.9}\scalebox{0.95}{$\{2,\mathcal{R}\}_{{\rm M}1}$}&\scalebox{0.9}{$-5.97 + 9.69\,\xi_*+ 0.0895\,\xi_*^2$}&\scalebox{0.9}{$2.30 +0.518\, \xi_* + 0.0117\, \xi_* ^2$}&\scalebox{0.9}{$1.10 -0.134\, \xi_* + 0.0087\, \xi_* ^2$} \\\hline
\cellcolor[gray]{0.9}\scalebox{0.95}{$\{2,-\}_{{\rm M}1}$} &\scalebox{0.9}{$-7.50 +9.69\, \xi_* + 0.0920\, \xi_* ^2$}&\scalebox{0.9}{$3.84 + 0.652\,\xi_* + 0.0291\, \xi_*^2$}&\scalebox{0.9}{$1.06 -0.147\,\xi_* + 0.0094\,\xi_* ^2 $}\\\hline
\hline
\hline
\cellcolor[gray]{0.9}\scalebox{0.95}{$\{2,\mathcal{R}\}_{{\rm M}2}$}&\scalebox{0.9}{$ -15.13 + 10.09\,\xi_*+ 0.0389\,\xi_*^2$}&\scalebox{0.9}{$6.63 -0.403\, \xi_* + 0.0856\, \xi_* ^2$}&\scalebox{0.9}{$0.89 -0.101\, \xi_* + 0.0066\, \xi_* ^2$} \\\hline
\cellcolor[gray]{0.9}\scalebox{0.95}{$\{2,-\}_{{\rm M}2}$}&\scalebox{0.9}{$-14.78 +9.91\, \xi_* + 0.0487\, \xi_* ^2$}&\scalebox{0.9}{$7.78 - 0.166\,\xi_* + 0.0992\, \xi_*^2$}&\scalebox{0.9}{$0.83 -0.110\,\xi_* + 0.0070\,\xi_* ^2$}\\\hline
\hline
\end{tabular}
\caption{\label{tab:fit1} Fitting formulas for the height $f^{c}_{2,j}$, location $x^c_{2,j}$ and width $\sgm_{2,j}$ of peak of the scale dependent enhancement functions $f_{2,j}$ in \eqref{SC} for $\delta = 0.3$ and $3\leq \xi_*\leq 6.5$ in Model 1 (top) and Model 2 (bottom).}
\end{center}			
\end{table}
Here, we provide some details regarding the 2-pt power spectra in the rolling spectator axion-gauge field model. Using the definition \eqref{defr} of the tensor-to-scalar ratio with eqs. \eqref{psv} and \eqref{SC}, the full expression for $r$ can be written as

\beq\label{rapp}
r(k) \simeq 16 \epsilon_\phi \left[ \fr{1+ \fr{\epsilon_\phi}{16}\, \mathcal{P}^{(\rm v)}_\mathcal{R}(k)\, f_{2,-}\left(\xi_*, \fr{k}{k_*},\delta \right)}{1+ \epsilon_\phi^2\, \mathcal{P}^{(\rm v)}_\mathcal{R}(k)\, f_{2,\mathcal{R}}\left(\xi_*,\fr{k}{k_*},\delta\right)} \right], 
\eeq
where $\sum_\lambda \mathcal{P}^{(\rm v)}_\lambda \equiv  \mathcal{P}^{(\rm v)}_h = 16\epsilon_\phi \mathcal{P}^{(\rm v)}_{\mathcal{R}}$ is the total tensor vacuum power spectrum and we have neglected the subdominant positive helicity mode of sourced fluctuations $f_{2,+} \ll f_{2,-}$. In the parametrization provided in \eqref{rapp}, the second terms in the numerator and denominator give the ratio between the sourced and vacuum power spectrum for tensor/scalar fluctuations respectively: 
\beq\label{Rts}
R_{t} \equiv \fr{\epsilon_\phi}{16}\, \mathcal{P}^{(\rm v)}_\mathcal{R}(k)\, f_{2,-}\left(\xi_*,\fr{k}{k_*},\delta\right), ~~~ R_s \equiv  \epsilon_\phi^2\, \mathcal{P}^{(\rm v)}_\mathcal{R}(k)\, f_{2,\mathcal{R}}\left(\xi_*,\fr{k}{k_*},\delta\right).
\eeq 
To evaluate the expressions \eqref{rapp} and \eqref{Rts}, in Table \ref{tab:fit1}, we provide fitting formulas for the height, width and the position of the peak of $f_{2,j}$ in \eqref{SC} using the exact expressions of these functions that appeared in the Appendixes of \cite{Namba:2015gja,Ozsoy:2020ccy}. We can then use the fitting formulas in Table \ref{tab:fit1} to evaluate the tensor to scalar ratio \eqref{rapp} and the ratios of the sourced to vacuum power spectra in \eqref{Rts} at the peak scale of the GW signal $k = k_* \,x^c_{2,-}$. The constant $r$ curves together with the various $R_t$ and $R_s$ obtained in this way are shown in the $\epsilon_\phi-\xi_*$ plane in Figure \ref{fig:rpeak}. When the sourced contribution dominates $R_t \gg 1$, we found that tensor to scalar ratio \eqref{rapp} evaluated at its peak $k = k_{\rm p} = k_* x^{c}_{2,-}$ can be well approximated by the formula
\beq\label{rpeak}
r_{\rm p}\simeq \epsilon_\phi^2
 \begin{dcases} 
       e^{10.61 (\xi_* - 2.81)}\quad({\rm Model}\,\,1)\\
       e^{10.40(\xi_*-3.46)}\quad({\rm Model}\,\, 2),
   \end{dcases}
\eeq 
where we linearized the exponent of ($\ln (f^{c}_{2,-})$) in Table \ref{tab:fit1} within the range $3.5 \leq \xi_* \leq 6.5$.
\section{Shape analysis of SSS and TTT correlators}\label{AppBB}
In this appendix, we study shape of the scalar and tensor auto bispectrum in the non-compact axion monodromy model we described in the main text (See \eg \eqref{pots}). Similar to the mixed non-Gaussianity, the bispectrum can be factorized as \cite{Ozsoy:2020ccy}, 
\beq\label{3ptauto}
\mathcal{B}^{(s)}_{\,\,\,j}(\vec{k}_1,\vec{k}_2,\vec{k}_3) =  \frac{\left[\epsilon_{\phi} \mathcal{P}_{\mathcal{R}}^{(v)}\right]^{3}}{(k_{1} k_{2} k_{3})^2}\, f^{(3)}_{\,\,\,j}\left(\xi_{*}, \delta,  x_{*}, x_{2}, x_{3}\right),
\eeq
\begin{figure}[t!]
\begin{center}
\includegraphics[scale=0.82]{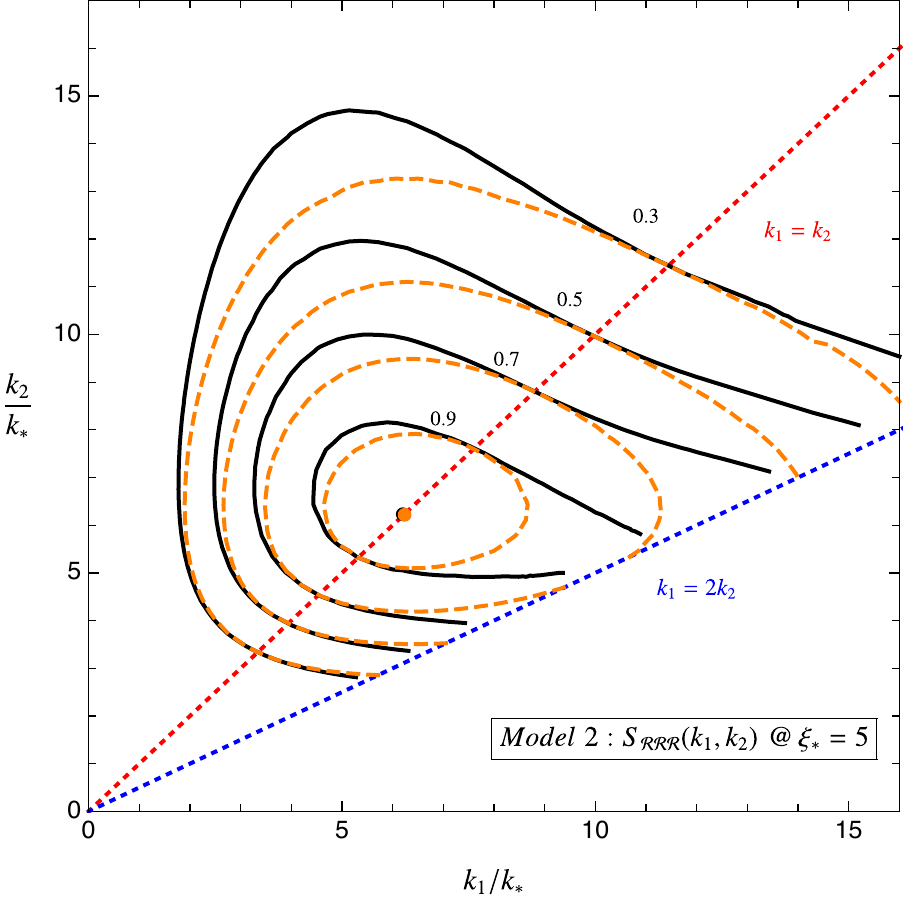}\,\,\,\,\includegraphics[scale=0.82]{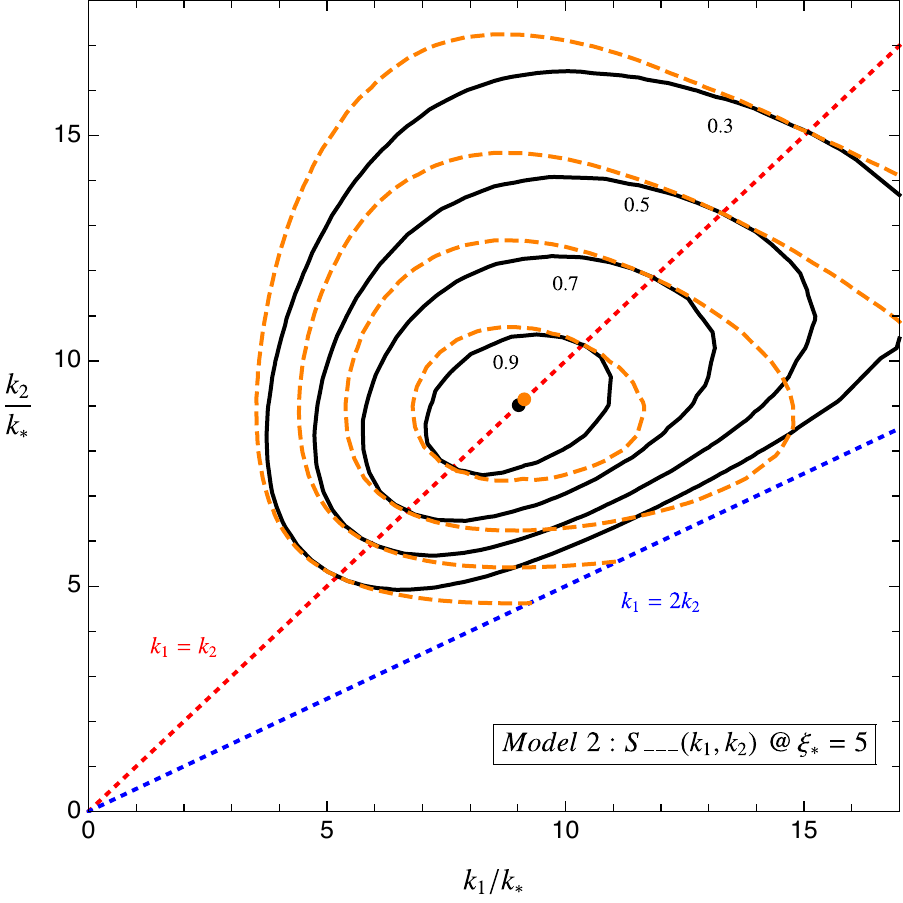}
\end{center}
\caption{Constant contour lines of the shape functions $S_{\mathcal{R}\mathcal{R}\mathcal{R}}$ and $S_{---}$, respectively of the $\langle\hat{\mathcal{R}}\hat{\mathcal{R}}\hat{\mathcal{R}}\rangle$ and $\langle\hat{h}_{-}\hat{h}_{-}\hat{h}_{-}\rangle$ bispectra in \eqref{sf} in the non-compact axion model (M2) for an isoceles triangle $k_2 = k_3$ and for $\{\xi_* = 5, \delta = 0.3 \}$. The black dots locate the triangle configuration at which bispectrum is maximum whereas the orange dot represent the location of the maximum obtained from the approximate expression \eqref{autobapp}. The dotted lines are drawn for reference to equilateral triangles $k_1 = k_2$ and folded triangles $k_1 = 2 k_2$. \label{fig:autob}}
\end{figure}
where $j =\{\mathcal{R}\mathcal{R}\mathcal{R}, ---\}$ and $k_1 = k, x_2 = k_2/k, x_3 = k_3 /k $. To analyze the shape, we use the definition of the shape function \eqref{sf} and utilize the explicit formulas derived in the Appendix B and C of \cite{Ozsoy:2020ccy} (See \eg eq. (B.19) and (C.15) in \cite{Ozsoy:2020ccy}). We then focus on isoceles triangles $x_2 = x_3$ to numerically evaluate these exact expressions on a grid of values in the $k_1 / k_*$ vs $k_2 /k_*$ plane and plot in Figure \ref{fig:autob} the constant contour lines (black solid lines) of $S_j$ that correspond to $0.9, 0.7, 0.5, 0.3$ of its maximal value (black dots) where $S_j(k_1,k_2) = 1$. We see that similar to the Model 1 studied in \cite{Namba:2015gja}, both bispectra is maximal on an equilateral triangles of scales $k_1 \simeq k_2 = k_3 \simeq \mathcal{O}(5-10) k_*$ that is approximately equal to the scales at which power spectra has a peak (See Table \ref{tab:fit1}). Motivated by this and the invariance of auto bispectra under the exchange of any pair external momenta, an approximate factorized form for the shape function $S \propto f^{(3)}_j$ is postulated in \cite{Namba:2015gja}:
\beq\label{autobapp}
f^{(3)}_{j}\left(k_{1}, k_{2}, k_{3}\right) \simeq\left[\frac{f^{(3)}_{j}\left(k_{1}, k_{1}, k_{1}\right)}{3 f_{2, j}\left(k_{1}\right)^{3 / 2}}+\frac{f^{(3)}_{j}\left(k_{2}, k_{2}, k_{2}\right)}{3 f_{2, j}\left(k_{2}\right)^{3 / 2}}+\frac{f^{(3)}_{j}\left(k_{3}, k_{3}, k_{3}\right)}{3 f_{2, j}\left(k_{3}\right)^{3/2}}\right] \prod^{3}_{i=1} f_{2, j}\left(k_{i}\right)^{1/2},
\eeq
where we omit the dependence of the sourced quantities $f_{2,j}$ and $f^{(3)}_j$ on $\xi_*$, $\delta$ and $k_*$ for the simplicity of the notation.
The accuracy of \eqref{autobapp} in capturing the actual shape of the bispectra is shown in Figure \ref{fig:autob}.  
 \section{Computations of the mixed bispectra}\label{AppC}
We present here our derivation of the STT and TSS bispectrum. For this purpose, we first note eq. \eqref{CPSapp} and the definitions of electric and magnetic fields in \eqref{EBF}, to write $\mathcal{R}$ as \cite{Namba:2015gja,Ozsoy:2020ccy},
\begin{align}\label{sR}
\nn \hat{\mathcal{R}}^{(\rm s)}(0,\vec{k}) &= \left(\fr{H}{\Mp}\right)^2 \fr{3\sqrt{2\pi^3}\xi_*}{8 k^4} \int \frac{\d^{3} p}{(2 \pi)^{3 / 2}} \,\epsilon^{-}_{i}(\vec{k}-\vec{p}) \epsilon^{-}_{i}(\vec{p})\,\, p^{1 / 4}\,|\vec{k}-\vec{p}|^{1 / 4}\left(p^{1/2}+ |\vec{k}-\vec{p}|^{1/2}\right) \\\nn
&\quad\quad\quad\quad\quad\quad\quad\quad\quad\quad\times  N_A\big(\xi_*, -|\vec{k}-\vec{p}|\tau_*,\delta\big)N_A\big(\xi_*, -p\tau_*,\delta\big)\, \hat{\mathcal{O}}_{-}(\vec{k}-\vec{p})\, \hat{\mathcal{O}}_{-}(\vec{p})\\
&\quad\quad\quad\quad\quad\quad\quad\quad\quad\quad \times  \mathcal{I}_{\mathcal{R}}\bigg[\xi_*,x_*,\delta,\sqrt{\fr{|\vec{k}-\vec{p}|}{k}} +\sqrt{\fr{p}{k}}\bigg].
\end{align}
In \eqref{sR}, $\mathcal{I}_\mathcal{R}$ includes a time integration over the gauge field mode functions \cite{Namba:2015gja,Ozsoy:2020ccy},
\begin{align}\label{Rs}
\nn \mathcal{I}_{\mathcal{R}}\bigg[\xi_{*}, x_{*}, \delta, Q\bigg] \equiv \int_{0}^{\infty} \frac{d x^{\prime}}{x^{\prime}} J_{3 / 2}\left(x^{\prime}\right) \sqrt{\frac{\epsilon_{\sigma}\left(x^{\prime}\right)}{\epsilon_{\sigma, *}}} \int_{x^{\prime}}^{\infty} d x^{\prime \prime} x^{\prime \prime 3 / 2} \exp \left[-E(x'')\sqrt{2\xi_{*}\,x''} \,Q\right] \\
\quad \times\left[J_{3 / 2}\left(x^{\prime}\right) Y_{3 / 2}\left(x^{\prime \prime}\right)-Y_{3 / 2}\left(x^{\prime}\right) J_{3 / 2}\left(x^{\prime \prime}\right)\right],
\end{align}
where $\sqrt{{\epsilon_{\sigma}(x^{\prime})}/{\epsilon_{\sigma, *}}} = {(1+\ln\left[(x_*/x' )^{\delta}\right]^2)^{-1}}$. 

Similarly, plugging the definitions \eqref{EBF} in \eqref{sh} and noting \eqref{ts}, $\hat{h}^{(\rm s)}_\lambda$ is given by \cite{Namba:2015gja,Ozsoy:2020ccy},
\begin{align}\label{hTs}
\nn\hat{h}^{(\rm s)}_\lambda (0, k) &\simeq \sqrt{\fr{2}{k^{7}}}\left(\fr{H}{\Mp}\right)^2  \int \frac{\d^{3} p}{(2 \pi)^{3 / 2}} \epsilon_{\lambda}\left[\vec{k},\vec{k}-\vec{p}, \vec{p}\right]\, p^{1 / 4}\,|\vec{k}-\vec{p}|^{1 / 4} N_A\big(\xi_*, -|\vec{k}-\vec{p}|\tau_*,\delta\big)\\
&\quad\quad\quad\quad\quad\quad\quad\quad \times  N_A\big(\xi_*, -p\tau_*,\delta\big)\mathcal{I}_h\bigg[\xi_*,x_*,\delta,\fr{|\vec{k}-\vec{p}|}{k},\fr{p}{k}\bigg]\, \hat{\mathcal{O}}_{-}(\vec{k}-\vec{p})\, \hat{\mathcal{O}}_{-}(\vec{p}),
\end{align}
where we defined the product of helicity vectors
\beq\label{epsl}
 \epsilon_{\lambda} \left[ \vec{k}, \vec{k}-\vec{p},\vec{p}  \right] \equiv \epsilon_i^{\lambda} (\vec{k})^{*}\,  \epsilon_i^{-} (\vec{k}-\vec{p})\,  \epsilon_j^{\lambda} (\vec{k})^{*} \, \epsilon_j^{-} (\vec{p})
 \eeq
 and 
\beq\label{inth}
\mathcal{I}_h\bigg[\xi_{*}, x_{*}, \delta, \tilde{p}, \tilde{q}\bigg] \equiv \mathcal{I}^{(1)}_h\bigg[\xi_{*}, x_{*}, \delta, \sqrt{\tilde{p}}+\sqrt{\tilde{q}}\bigg]+\frac{\sqrt{\tilde{p}\, \tilde{q}}}{2} \mathcal{I}^{(2)}_h\bigg[\xi_{*}, x_{*}, \delta, \sqrt{\tilde{p}}+\sqrt{\tilde{q}}\bigg]
\eeq
where $\mathcal{I}^{(1)}_h$ and $\mathcal{I}^{(2)}_h$ contains temporal integration of the gauge field sources \cite{Namba:2015gja,Ozsoy:2020ccy},
\begin{align}\label{inth2}
\nn&\mathcal{I}^{(1)}_{h}\bigg[\xi_{*}, x_{*}, \delta, Q\bigg] \equiv \int_{0}^{\infty} \d x^{\prime}\left(x^{\prime} \cos x^{\prime}-\sin x^{\prime}\right) \sqrt{\frac{\xi\left(x^{\prime}\right)}{x^{\prime}}} \exp \left[-E(x') \sqrt{2\xi_{*}\,x'} \,Q\right]\\
&\mathcal{I}^{(2)}_{h}\bigg[\xi_{*}, x_{*}, \delta, Q\bigg] \equiv \int_{0}^{\infty} \d x^{\prime}\left(x^{\prime} \cos x^{\prime}-\sin x^{\prime}\right) \sqrt{\frac{x^{\prime}}{\xi\left(x^{\prime}\right)}} \exp \left[-E(x') \sqrt{2\xi_{*}\,x'} \,Q\right].
\end{align}
Note that for the sourced scalar and tensor perturbations, the dependence of $\xi(x)$ and $E(x)$ on the axion potential in the spectator sector are provided in \eqref{Joep} and \eqref{Q} respectively.
\smallskip

\noindent{\bf STT and TSS Bispectrum.} Noting the definitions of the mixed bispectra in \eqref{DBS}, we are ready to calculate $\langle\mathcal{R}h_{\lambda} h_{\lambda}\rangle$ and $\langle h_{\lambda}\mathcal{R} \mathcal{R}\rangle$ using $\hat{h}_\lambda$ \eqref{hTs} and $\hat{\mathcal{R}}$ \eqref{sR}. For this purpose, we employ the Wick's theorem to compute the products of $\hat{\mathcal{O}}_-$ operators to obtain the following form for the mixed bispectra
\begin{align}\label{BS}
\nn \mathcal{B}^{(\rm s)}_{\mathcal{R}{\lambda}{\lambda}}(\vec{k}_1,\vec{k}_2,\vec{k}_3) &\simeq \frac{\left[\epsilon_{\phi} \mathcal{P}_{\mathcal{R}}^{(v)}\right]^{3}}{(k_{1} k_{2} k_{3})^2}\, f^{(3)}_{\mathcal{R}\lambda\lambda}\left(\xi_{*},  \delta, x_{*}, x_{2}, x_{3}\right),\\
\mathcal{B}^{(\rm s)}_{{\lambda}\mathcal{R}\mathcal{R}}(\vec{k}_1,\vec{k}_2,\vec{k}_3) &\simeq \frac{\left[\epsilon_{\phi} \mathcal{P}_{\mathcal{R}}^{(v)}\right]^{3}}{(k_{1} k_{2} k_{3})^2}\, f^{(3)}_{\lambda\mathcal{R}\mathcal{R}}\left(\xi_{*},  \delta, x_{*}, x_{2}, x_{3}\right),
\end{align}
where we used the standard vacuum contribution to the scalar power spectrum \eqref{psv} to replace factors of $(H/\Mp)^6$ and we have fixed $k_1 = k$, $k_2 = x_2 k$ and $k_3 = x_3 k$. 
As indicated by the diagrams in Figure \ref{fig:diag}, mixed correlators arise as a result of a ``loop" computation over the internal momentum that labels gauge field modes. Defining the rescaled internal momentum as $\vec{\tilde{p}} = \vec{p}/k$, we found that dimensionless $f^{(3)}$ functions that parametrize this computation are given by
\begin{align}\label{f3RMM}
\nn f^{(3)}_{\mathcal{R}\lambda\lambda}&= \frac{192 \pi^3 \xi_{*}}{(x_{2} x_{3})^{3/2}} \int \d^3 \tilde{p} \,\, \epsilon_{(\mathcal{R}\lambda\lambda)}\left[\vec{\tilde{p}}, \hat{k}_{1}, \hat{k}_{2},\hat{k}_{3}\right]\sqrt{\tilde{p}\left|\hat{k}_{1}-\vec{\tilde{p}}\right|\left|\vec{\tilde{p}}+x_{2} \hat{k}_{2}\right|} \left(\sqrt{\tilde{p}}+\sqrt{\left|\hat{k}_{1}-\vec{\tilde{p}}\right|}\right)\\\nn&\quad\quad\quad\quad\quad\quad\times N_A^{2}\left(\xi_{*}, \tilde{p} x_{*}, \delta\right)  N_A^{2}\left(\xi_{*},\left|\hat{k}_{1}-\vec{\tilde{p}}\right| x_{*}, \delta\right)N_A^{2}\left(\xi_{*},\left|\vec{\tilde{p}}+x_{2} \hat{k}_{2}\right| x_{*}, \delta\right)\\\nn
&\quad\quad\quad\quad\quad\quad\times \mathcal{I}_{\mathcal{R}}\left[\xi_{*}, x_{*}, \delta, \sqrt{\tilde{p}}+\sqrt{\left|\hat{k}_{1}-\vec{\tilde{p}}\right|}\right] \mathcal{I}_{h}\left[\xi_{*}, x_{2} x_{*}, \delta,\frac{\tilde{p}}{x_2},\frac{\left|\vec{\tilde{p}}+x_{2} \hat{k}_{2}\right|}{x_2}\right]\\
&\quad\quad\quad\quad\quad\quad\times \mathcal{I}_{h}\left[\xi_{*}, x_{3} x_{*}, \delta, \frac{\left|\hat{k}_{1}-\vec{\tilde{p}}\right|}{x_{3}},\frac{\left|\vec{\tilde{p}}+x_{2} \hat{k}_{2}\right|}{x_3}\right],
\end{align}
and
\begin{align}\label{f3MRR}
\nn f^{(3)}_{\lambda\mathcal{R}\mathcal{R}}&= \frac{72 \pi^{9/2} \xi_{*}^2}{(x_{2} x_{3})^{2}} \int \d^3 \tilde{p}\, \, \epsilon_{(\lambda\mathcal{R}\mathcal{R})}\left[\vec{\tilde{p}}, \hat{k}_{1}, \hat{k}_{2},\hat{k}_{3}\right]\left(\sqrt{\tilde{p}}+\sqrt{\left|\vec{\tilde{p}}+x_{2} \hat{k}_{2}\right|}\right)\left(\sqrt{\left|\vec{\tilde{p}}+x_{2} \hat{k}_{2}\right|}+\sqrt{\left|\hat{k}_{1}-\vec{\tilde{p}}\right|}\right) \\\nn&\quad\quad\times \sqrt{\tilde{p}\left|\hat{k}_{1}-\vec{\tilde{p}}\right|\left|\vec{\tilde{p}}+x_{2} \hat{k}_{2}\right|}N_A^{2}\left(\xi_{*}, \tilde{p} x_{*}, \delta\right)  N_A^{2}\left(\xi_{*},\left|\hat{k}_{1}-\vec{\tilde{p}}\right| x_{*}, \delta\right)N_A^{2}\left(\xi_{*},\left|\vec{\tilde{p}}+x_{2} \hat{k}_{2}\right| x_{*}, \delta\right)\\\nn
&\quad\quad\quad\times \mathcal{I}_{h}\left[\xi_{*}, x_{*}, \delta, \tilde{p},\left|\hat{k}_{1}-\vec{\tilde{p}}\right|\right] \mathcal{I}_{\mathcal{R}}\left[\xi_{*}, x_{2} x_{*}, \delta,\frac{\sqrt{\tilde{p}} + \sqrt{\left|\vec{\tilde{p}}+x_{2} \hat{k}_{2}\right|}}{\sqrt{x_2}}\right] \\
&\quad\quad\quad\quad\quad\quad\times \mathcal{I}_{\mathcal{R}}\left[\xi_{*}, x_{3} x_{*}, \delta, \frac{\sqrt{\left|\hat{k}_1-\vec{\tilde{p}}\right|} + \sqrt{\left|\vec{\tilde{p}}+x_{2} \hat{k}_{2}\right|}}{\sqrt{x_3}}\right].
\end{align}
For the numerical evaluation of the integrals, keeping the definition \eqref{epsl} in mind, we note the product of helicity vectors in \eqref{f3RMM} and \eqref{f3MRR} as
\begin{align}\label{pohv}
\nn \epsilon_{(\mathcal{R}\lambda\lambda)}
&\equiv \epsilon^{-}_i(\hat{k}_{1}-\vec{\tilde{p}}) \epsilon^{-}_i(\vec{\tilde{p}})\, \epsilon_{\lambda} \left[ \hat{k}_2, -\vec{\tilde{p}},\vec{\tilde{p}}+x_2\hat{k}_2  \right]  \epsilon_{\lambda} \left[ \hat{k}_3, -(\hat{k}_1-\vec{\tilde{p}}), -(\vec{\tilde{p}}+x_2\hat{k}_2) \right] ,\\
 \epsilon_{(\lambda\mathcal{R}\mathcal{R})}
 & \equiv \epsilon_{\lambda} \left[ \hat{k}_1, \hat{k}_1-\vec{\tilde{p}},\vec{\tilde{p}} \right]
\epsilon^{-}_i(-\vec{\tilde{p}}) \epsilon^{-}_i(\vec{\tilde{p}}+x_2 \hat{k}_2)\,\epsilon^{-}_j(-(\hat{k}_1-\vec{\tilde{p}})) \epsilon^{-}_j(-(\vec{\tilde{p}}+x_2 \hat{k}_2)).
 \end{align}
Finally, we align $\vec{k}_1$ along the x axis, $\vec{k}_{1}=k\, (1,0,0)$ and express $\vec{k}_2$ and $\vec{k}_3$ in terms of $x_2$ and $x_3$,
\begin{align}\label{emconfig}
\nn \vec{k}_{2}&= k\, x_2 \left( \fr{-1-x_{2}^{2}+x_{3}^{2}}{2 x_2},\fr{\sqrt{-\left(1-x_{2}+x_{3}\right)\left(1+x_{2}-x_{3}\right)\left(1-x_{2}-x_{3}\right)\left(1+x_{2}+x_{3}\right)}}{2 x_2},0\right),\\
\vec{k}_{3}&=k\, x_3 \left(\fr{-1+x_{2}^{2}-x_{3}^{2}}{2 x_3},-\fr{\sqrt{-\left(1-x_{2}+x_{3}\right)\left(1+x_{2}-x_{3}\right)\left(1-x_{2}-x_{3}\right)\left(1+x_{2}+x_{3}\right)}}{2 x_3},0\right).
\end{align}
and define the polarization vector for a given momentum $\vec{q}$ in terms of its components as
\beq\label{epsdef}
\epsilon^{\lambda}(\vec{q})=\frac{1}{\sqrt{2}}\left(\frac{q_{x} q_{z} - i \lambda q_{y} |\vec{q}|}{ |\vec{q}|~\sqrt{q_{x}^{2}+q_{y}^{2}}  },\frac{q_{y} q_{z} + i \lambda q_{x} |\vec{q}| }{ |\vec{q}|~\sqrt{q_{x}^{2}+q_{y}^{2}}},-\frac{\sqrt{q_{x}^{2}+q_{y}^{2}}}{|\vec{q}|} \right).
\eeq
The shape and scale dependence of $f^{(3)}$ functions can then be obtained numerically by fixing the background parameters $\{\xi_*, \delta\}$ that parametrize the efficiency of the particle production process in the gauge field sector.

\noindent{\bf Properties of the mixed bispectra.} Let us verify some basic properties of the mixed bispectra \eqref{BS} we derived here. To show the invariance of $\mathcal{B}_j$ under the exchange of $\vec{k}_2 \leftrightarrow \vec{k}_3$, we first replace $\vec{k}_2 \to \vec{k}_3$ and $\vec{k}_3 \to \vec{k}_2$ on both sides of the expressions in \eqref{BS} using \eqref{f3MRR}, \eqref{f3RMM} and \eqref{pohv}. Then changing the integration variable $\vec{p} \to -\vec{p} + \vec{k}_1$ and noting $\sum_i \vec{k}_i = 0$, it is easy confirm that the resulting expressions is equivalent to \eqref{f3MRR} and \eqref{f3RMM}. To prove that $\mathcal{B}_j$ is real, we first use the reality of $h_{ij}$ and $\mathcal{R}$ in the configuration space, which implies ${h}_\lambda(\vec{k}) = {h}^*_\lambda(-\vec{k})$ and $\mathcal{R}(\vec{k}) = \mathcal{R}^*(-\vec{k})$. Using the last two identities, then $\mathcal{B}_j(\vec{k}_1,\vec{k}_2,\vec{k}_3) = \mathcal{B}_j^*(-\vec{k}_1,-\vec{k}_2,-\vec{k}_3) $ follows immediately. Finally, focusing on the latter quantity, we perform a $180^\circ$ rotation around the axis $\perp$ to plane defined by $\sum_i \vec{k}_i = 0$ to change the orientation of the external momenta $-\vec{k}_i \to \vec{k}_i$ in its arguments and note the invariance of the bispectrum under this action due to isotropy of the background which together implies $\mathcal{B}_j^*(-\vec{k}_1,-\vec{k}_2,-\vec{k}_3) = \mathcal{B}_j^*(\vec{k}_1,\vec{k}_2,\vec{k}_3) $ and hence $ \mathcal{B}_j(\vec{k}_1,\vec{k}_2,\vec{k}_3) =  \mathcal{B}_j^*(\vec{k}_1,\vec{k}_2,\vec{k}_3)$. From these arguments, we also infer the following relation $\mathcal{B}_j(-\vec{k}_1,-\vec{k}_2,-\vec{k}_3) = \mathcal{B}_j(\vec{k}_1,\vec{k}_2,\vec{k}_3)$.
\subsection{Peak structure of $f^{(3)}_{-\mathcal{R}\mathcal{R}}$ vs $f^{(3)}_{\mathcal{R}--}$ at the equilateral configuration}\label{AppC1}
To understand the peak structure of the mixed $-\mathcal{R}\mathcal{R}$ and $\mathcal{R}--$ correlators, we focus our attention to the integrands of eqs. \eqref{f3RMM} and \eqref{f3MRR} at the equilateral configuration $x_2 = x_3 = 1$. The integrands depend on the magnitude of momentum $\tilde{p}$ running in the loop (see Figure \ref{fig:diag}), and its orientation --parametrized by the polar $\theta$ and azimuthal angle $\phi$-- with respect to the plane ($x$-$y$) where external momenta $\vec{k}_i$ lives (see eq. \eqref{emconfig}). At fixed $\delta$, their structure can be schematically written as 
\beq\label{intf}
If^{(3)}_j(\xi_*, x_*, \tilde{p},\theta,\phi) \propto \underbrace{\epsilon_{(j)}[\vec{\tilde{p}}, \vec{k}_1,\vec{k}_2,\vec{k}_3]}_{\textrm{(a)}}\,\underbrace{\mathcal{I}_h(\dots)\mathcal{I}_{h/\mathcal{R}}(\dots) \mathcal{I}_{\mathcal{R}}(\dots)}_{\textrm{(b)}}\, \underbrace{N^c_A[\xi_*]^6\,\tilde{p}^n\, e^{-\frac{3}{\sigma_A[\xi_{*}]^2} \ln ^{2}\left(\frac{ \tilde{p}\, x_*}{q^c_A[\xi_*]}\right)}}_{\textrm{(c)}},
\eeq
where $n = \{4,9/2\}$ for the $\mathcal{R}--$  and  $-\mathcal{R}\mathcal{R}$ correlator respectively. We discuss physical implications of the parts contributing to the integrand \eqref{intf} below. 
\begin{figure}[t!]
\begin{center}
\includegraphics[scale=0.85]{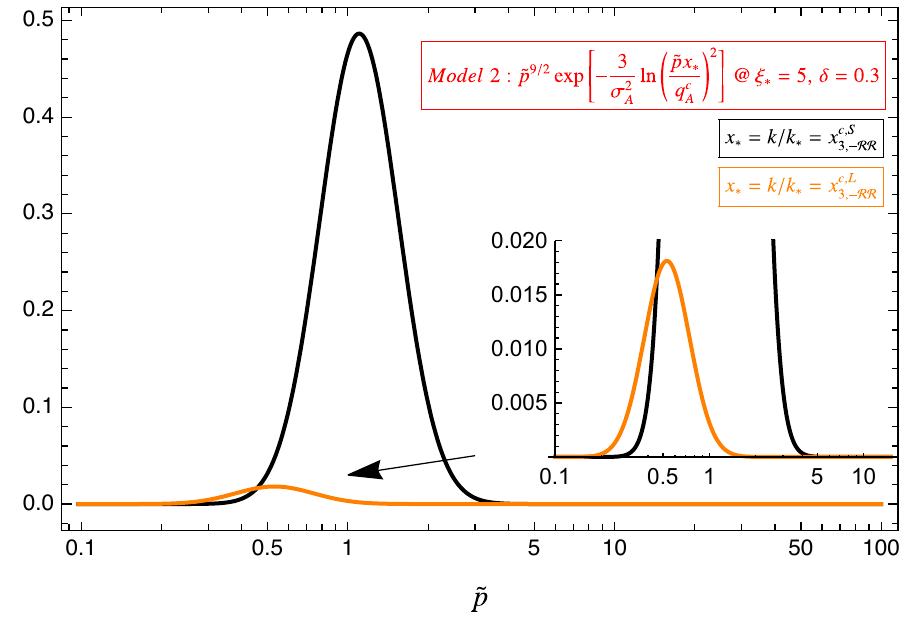}\includegraphics[scale=0.86]{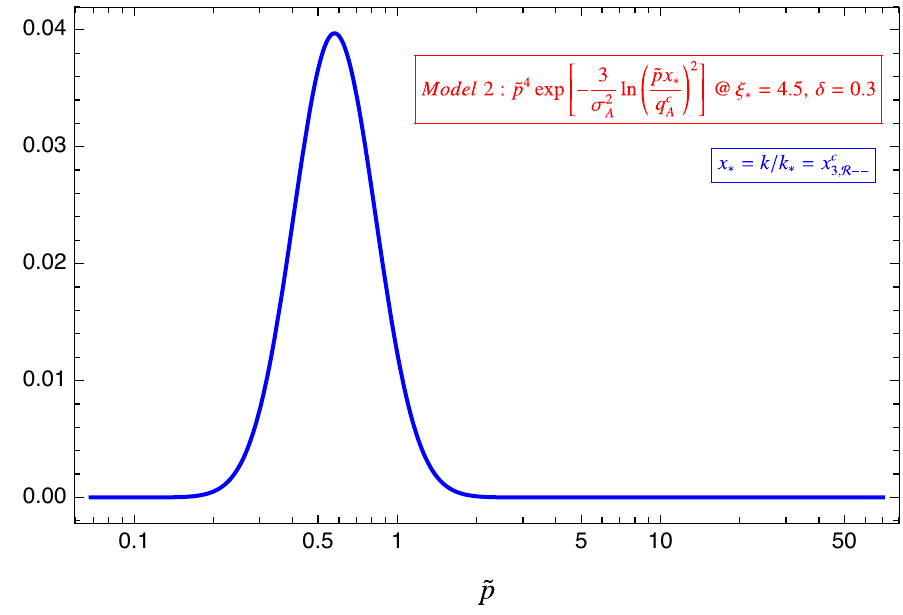}
\end{center}
\caption{Dependence of the amplified gauge field modes on the magnitude of loop momentum $\tilde{p}$ for $-\mathcal{R}\mathcal{R}$ (Left) and $\mathcal{R}--$ (Right). Both plots are normalized with $N^c_A[\xi_*]^6$: see \eg (c) in \eqref{intf}. \label{fig:c}}
\end{figure}
\begin{figure}[t!]
\begin{center}
\includegraphics[scale=0.85]{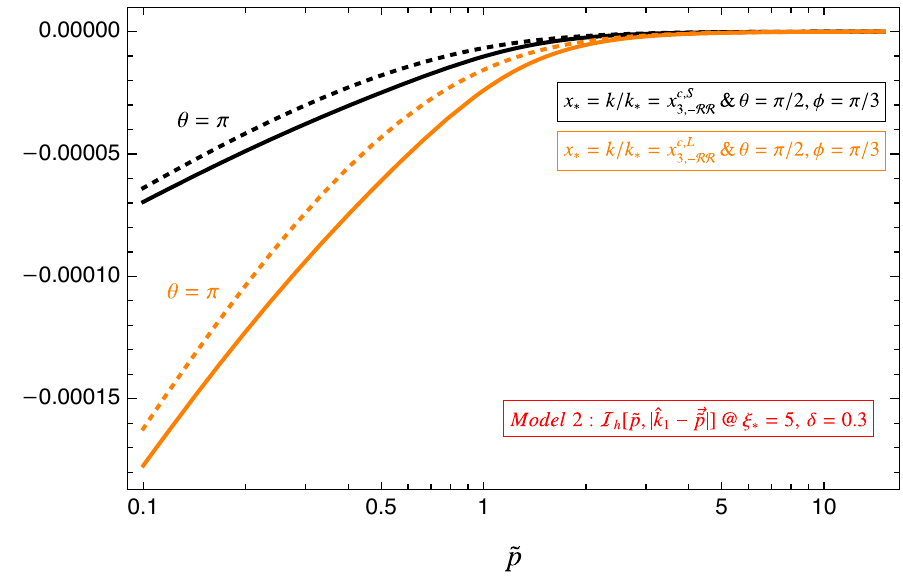}\includegraphics[scale=0.86]{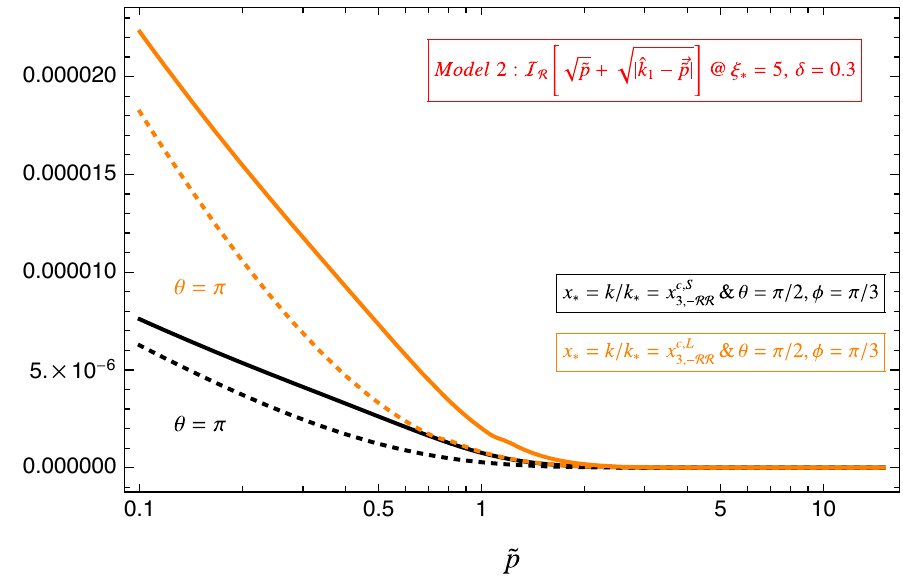}
\end{center}
\caption{Dependence of the $\mathcal{I}_{h/\mathcal{R}}$ (eqs. \eqref{inth}, \eqref{inth2} and \eqref{Rs}) on the magnitude of loop momentum $\tilde{p}$.\label{fig:b}}
\end{figure}
\begin{figure}[t!]
\begin{center}
\includegraphics[scale=0.85]{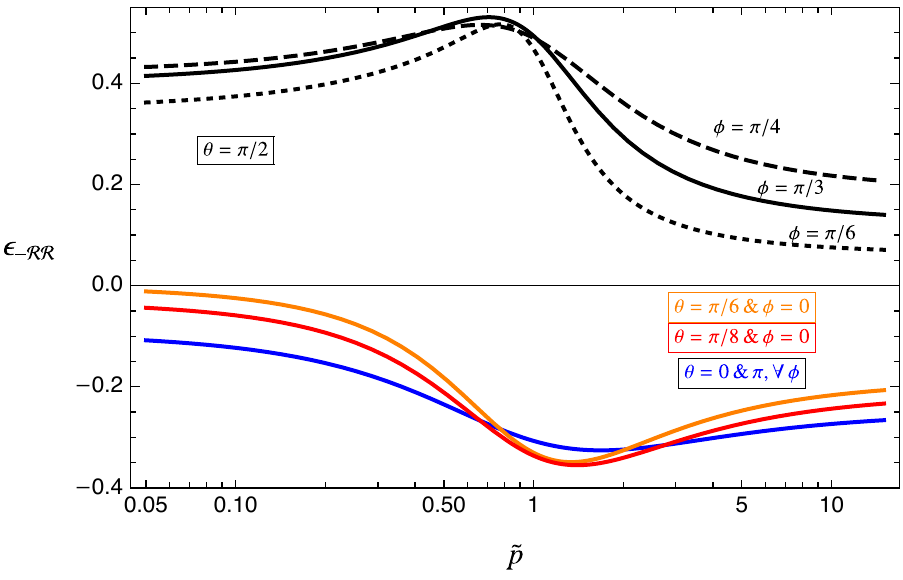}\includegraphics[scale=0.86]{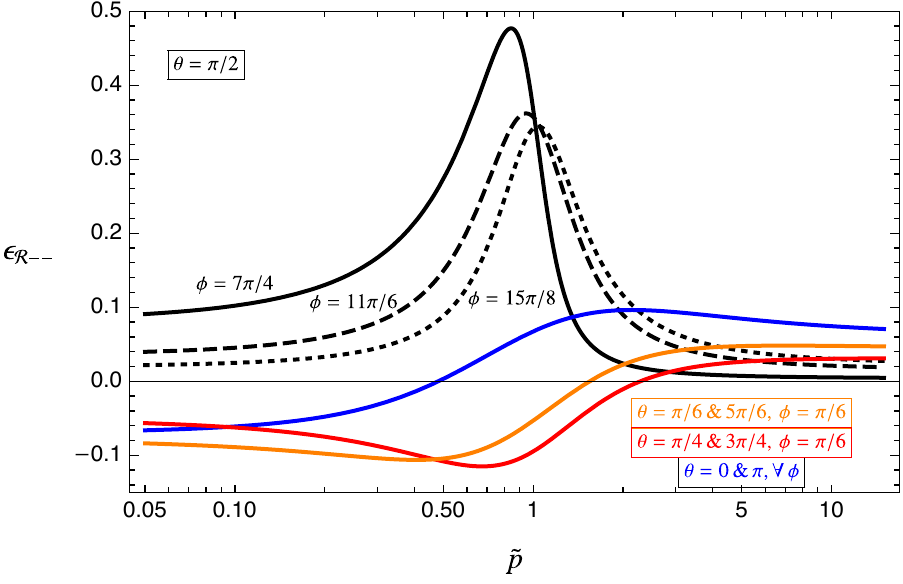}
\end{center}
\caption{The product of helicity vectors defined in \eqref{pohv} for $x_2=x_3 =1$ as a function of $\tilde{p}$ and for different loop momentum orientation with respect to the plane of external momenta $\vec{k}_i$. In both panels, the maximal positive contributions to the product arise when the loop momentum lies in the $x-y$ plane, namely when $\theta = \pi/2$.\label{fig:a}}
\end{figure}

\begin{itemize}
\item {\bf (c):} These terms identify the scale dependent amplitudes of gauge field mode functions in \eqref{Nform} and the manifestly $\tilde{\tilde{p}}$ dependent terms using the second and first line of the loop integrals in \eqref{f3RMM} and \eqref{f3MRR}. Physically, they characterize the scale dependent $x_* = k_i/k_* = k/k_*$ enhancement of the gauge modes running in the internal lines whose overall amplitude is dictated by the normalization factors $N^c_A[\xi_*]^6$ (see Table \ref{tab:fit0}). Notice that, in writing these terms, we ignored their $\theta$ and $\phi$ dependence in the $\tilde{p}^n \exp[\dots]$ part as the orientation of the loop momentum w.r.t to the plane of external momenta does have a little impact on the overall amplitude of gauge field modes compared to $\tilde{p}$. 
An essential feature of the terms labeled by (c) is that they acquire a peak (with an amplitude set by $N^c_A [\xi_*]^6$ factor) located at 
\beq\label{ppeak}
\tilde{p}_{\rm peak}\, x_* = e^{\fr{n \sigma_A[\xi_*]^2}{6}}\, q^c_A[\xi_{*}],
\eeq
which is important for understanding the scale dependence of the mixed correlators. In particular, \eqref{ppeak} implies that for larger (smaller) $x_* = k/k_*$, the loop integrals that characterize the correlators will have support around smaller (larger) values of $\tilde{p}_{\rm peak}$ because for $\tilde{p} > \tilde{p}_{\rm peak}$ or $\tilde{p} < \tilde{p}_{\rm peak}$, the terms labeled by (c) decay away quickly due to their exponential dependence. We illustrate these facts in Figure \ref{fig:c} where we plot  the terms labeled by (c) as a function of the magnitude of loop momentum $\tilde{p}$ for both correlators we focus and for different $k/k_*$.
\item {\bf (b):} The product of integrals $\mathcal{I}_{h/\mathcal{R}}$ in \eqref{intf} captures the propagation of the amplified gauge modes from the internal lines to the external lines characterized by the late time curvature $\mathcal{R}$ or tensor perturbation $h$ through the vertices shown in Figure \ref{fig:diag}. For the purpose of understanding peak structure of mixed correlators we are interested in, we plot them in Figure \ref{fig:b} in terms of $\tilde{p}$ for different $x_*$ and loop momentum configurations. We observe that for the range of $\tilde{p}$ values where the gauge field modes have appreciable contribution to the correlators (see Figure \ref{fig:c}) propagation effects associated with tensors are always negative $\mathcal{I}_h < 0$ whereas for the curvature perturbation, the same quantity is strictly positive $\mathcal{I}_\mathcal{R}> 0$. We found that this conclusion holds irrespective of the choice of loop momentum configurations parametrized by the polar $\theta$ and azimuthal angle $\phi$. 
\item {\bf (a):} These terms represent the scalar product of polarization vectors defined by \eqref{pohv} and \eqref{epsl} in the equilateral configuration $x_2 = x_3 =1$, and serve the purpose of helicity conservation at each vertex. Their behavior with respect to the magnitude of the loop momentum $\tilde{p}$ is crucial in understanding double peak vs single peak structure of the $-\mathcal{R}\mathcal{R}$ and $\mathcal{R}--$ correlators as we explain below. For orientations of loop momentum that leads to maximal results, we show the behavior of $\epsilon_{(-\mathcal{R}\mathcal{R})}[\vec{\tilde{p}}, \vec{k}_1,\vec{k}_2,\vec{k}_3]$ and $\epsilon_{(\mathcal{R}--)}[\vec{\tilde{p}}, \vec{k}_1,\vec{k}_2,\vec{k}_3]$ as a function of $\tilde{p}$ in Figure \ref{fig:a}. 
\end{itemize}

From the left panel of Figure \ref{fig:a}, we see that $\epsilon_{(-\mathcal{R}\mathcal{R})}$ has a significant negative support for loop momentum configurations that does not lie in the $x$-$y$ plane ($\theta \neq \pi/2$) in the $\tilde{p}\gtrsim 1$ regime. In this region, the integrand \eqref{intf} of the $-\mathcal{R}\mathcal{R}$ correlator \eqref{f3MRR} have significant support from the amplified gauge field mode functions at small $k/k_*$ (black curve in the left panel of Figure \ref{fig:c}) and integrating it over such loop momentum configurations leads to a positive peak at small $k/k_* = x^{c,S}_{3,-\mathcal{R}\mathcal{R}}$, recalling the overall negative sign of propagation effects $\mathcal{I}_{h} \mathcal{I}_{\mathcal{R}}\mathcal{I}_\mathcal{R}< 0$. On the other hand, for loop momentum that lives in the same plane with the external momenta ($\theta = \pi/2$), the product of polarization vectors have a positive support in the  $\tilde{p} \lesssim 1$ region. In this regime, the integrand  \eqref{intf} does still have support from the peak of amplified gauge field modes at larger $k/k_*= x^{c,L}_{3,-\mathcal{R}\mathcal{R}}$ (orange curve in the left panel of  Figure \ref{fig:c}). Therefore, $-\mathcal{R}\mathcal{R}$ correlator  obtains a second peak occuring in the negative direction due to the overall negative contributions arise from the propagation effects $\mathcal{I}_{h} \mathcal{I}_{\mathcal{R}}\mathcal{I}_\mathcal{R}< 0$. 

For the $\mathcal{R}--$ correlator, setting the product of polarization vectors aside, the integrand \eqref{intf} has an overall positive sign due to propagation effects $ \mathcal{I}_{\mathcal{R}}\mathcal{I}_h\mathcal{I}_h > 0$. More importantly, contrary to the case of  $-\mathcal{R}\mathcal{R}$ correlator, the range of loop momenta $\tilde{p}\lesssim 1$ where the amplified gauge field modes can contribute to the integrand (see the right panel in \ref{fig:c}) overlaps with the range where the product of helicity vectors takes its maximal values which is positive for $\theta = \pi/2$ as can be seen from Figure \ref{fig:a}. Integrating \eqref{intf} over such configurations therefore leads to a single peak for the $\mathcal{R}--$ correlator \eqref{f3RMM} occurring in the positive direction as the dominant support for $\epsilon_{(\mathcal{R}--)}$ is positive in this regime.

Considering that the $-\mathcal{R}\mathcal{R}$ and $\mathcal{R}--$ correlators differ from each other by an external scalar/ tensor state ($\mathcal{R}/h$) and comparing the left/right panel of Figure \ref{fig:a}, we can physically make sense of these results. In particular, for large enough transverse momentum $\tilde{p} \gtrsim 1$, conservation of angular momentum allows two internal photons to generate an external scalar perturbation $\mathcal{R}$ even if the latter lies in a plane different than the internal photons ($\theta \neq \pi/2$). In this way, one can generate soft $\mathcal{R}$'s to induce sizeable correlations between external states of  $-\mathcal{R}\mathcal{R}$ in the form of an early peak located at $k_i/k_* = k/k_* = x^{c,S}_{3,-\mathcal{R}\mathcal{R}} < x^{c,L}_{3,-\mathcal{R}\mathcal{R}}$ (See Figure \ref{fig:f3fit}). On the other hand, for soft internal momenta $\tilde{p} \lesssim 1$, internal photon states can still induce sizeable correlations between the external states of  $-\mathcal{R}\mathcal{R}$ correlator as far as the external momentum $\vec{k}_i$ lies in the same plane with the loop momentum $\vec{\tilde{p}}$ ($\theta=\pi/2$). Since the loop momentum does not leak beyond the plane of external momenta in this case, sizeable $-\mathcal{R}\mathcal{R}$ correlation can be induced at harder external momenta satisfying $k_i/k_* = k/k_* = x^{c,L}_{3,-\mathcal{R}\mathcal{R}} > x^{c,S}_{3,-\mathcal{R}\mathcal{R}}$, explaining the presence of a second peak in the $-\mathcal{R}\mathcal{R}$ correlator (See Figure \ref{fig:f3fit}). 

However, as can be seen from the right panel of Figure \ref{fig:a}, the same situation is more restrictive if the external state is a graviton. In this case, angular momentum conservation strictly prefers the production of an external graviton from two internal photons (preferably soft $\tilde{p}\lesssim 1$) that lie in the same plane and the resulting correlation between the external states of $\mathcal{R}--$ correlator is thus maximal at a single location parametrized by $k_i/k_* = x^c_{3,\mathcal{R}--}$.
In light of the discussion above, we conclude that the product of polarization vectors is the key quantity that determines the double peak vs single peak structure of mixed correlators. 
\section{Approximate factorized forms for the mixed bispectra}\label{AppD}
We now derive factorized approximate expressions for the STT \eqref{f3RMM} and TSS \eqref{f3MRR} bispectrum as a sum of terms given by the products of sourced signals $f_{2,j}(k_i)$ and $f^{(3)}_{j}(k_i,k_i,k_i)$ that contains only a single external momenta $k_i$.

{\bf $\bullet$ STT:}  Similar to the 3-pt auto correlators, we expect that the mixed spectra has a peaked structure so that we can utilize the 2-pt and 3-pt correlators (evaluated at the equilateral configuration) to describe it in a factorized form. Motivated by these considerations and the $\vec{k}_2 \leftrightarrow \vec{k}_3$ symmetry of the STT bispectrum, we start with the following ansatz: 
\begin{align}\label{f3rmmapp}
f^{(3,\rm app)}_{\mathcal{R}\lambda\lambda} &\simeq \mathcal{C}\left\{\frac{f^{(3)}_{\mathcal{R}\lambda\lambda}\left( s_1 k_{1}, s_1 k_{1}, s_1 k_{1}\right)}{f_{2, \mathcal{R}}\left(\bar{s}_1 k_{1}\right)^{1/2}f_{2,\lambda}\left(\tilde{s}_1 k_{1}\right)}+\left[\frac{f^{(3)}_{\mathcal{R}\lambda\lambda}\left(s_2 k_{2}, s_2 k_{2}, s_2 k_{2}\right)}{f_{2, \mathcal{R}}\left(\bar{s}_2 k_{2}\right)^{1 / 2}f_{2, \lambda}\left(\tilde{s}_2 k_{2}\right)}+ k_2 \to k_3\right]\right\} \\\nn 
&\quad\quad\quad\quad\quad\quad\quad\quad\quad\quad\quad\quad\quad\quad\quad\quad\quad\quad\quad\quad\quad\quad\quad \times \prod_{i=2,3} \left[ f_{2,\mathcal{R}}\left(\bar{s}_1 k_{1}\right) f_{2,\lambda}\left(\tilde{s}_2 k_{i}\right)\right]^{1/2},
\end{align}
where we introduced scaling factors for the external momenta in the sourced quantities $f_{2,j}$ and $f^{(3)}_j$ to be able to locate the maximum of the bispectra accurately in the $k_1 - k_2$ (recall that we focus on isosceles triangles $k_2 = k_3$). $\mathcal{C}$ is an overall coefficient that we will fix to re-produce the correct normalization of the exact bispectra as we describe below. 

To ensure that the approximate expression \eqref{f3rmmapp} describes the actual one accurately around its maximum, we can utilize the peak locations of the 2-pt functions $x^{c}_{2,j}$ (see Table \ref{tab:fit1}) and 3-pt functions evaluated at the equilateral configuration $x^c_{3,\mathcal{R}\lambda\lambda}$ (see Table \ref{tab:f3fit2}). In particular, since we know (by numerical evaluation) the triangle configuration at which the exact bispectra is maximal, say at $k_{1,2} = k^{\rm m}_{1,2}$, we can fix the scaling factors $s,\bar{s},\tilde{s}$ in \eqref{f3rmmapp} appropriately as $s_{1,2} = x^c_{3,j}/k^{\rm m}_{1,2}$, $\bar{s}_{1,2} = x^{c}_{2,\mathcal{R}}/k^{\rm m}_{1,2}$ and $\tilde{s}_{1,2} = x^c_{2,\lambda}/k^{\rm m}_{1,2}$ for a given set of model parameters $\xi_*$ and $\delta$. Considering the gaussian forms of the 2-pt \eqref{SC} and 3-pt mixed correlators (See \eg \eqref{f3fit}), the aforementioned choices of scaling factors provide a very accurate guess for the exact location of the maximum in the $k_1 - k_2$ plane. To fix the overall normalization $\mathcal{C}$, we then enforce the approximate expression \eqref{f3rmmapp} at its maximum to be equal to the maximum of the exact one derived from \eqref{f3RMM}, \ie $f^{3,\rm max}_{\mathcal{R}\lambda\lambda} = f^{(3)}_{\mathcal{R}\lambda\lambda}(k^{\rm m}_1,k^{\rm m}_2)$. 

{\bf $\bullet$ TSS:} For the TSS type correlators, following the same procedures above, we found that the following expression provide an accurate description of the exact bispectrum: 
\begin{align}\label{f3mrrapp}
f^{(3,\rm app)}_{\lambda\mathcal{R}\mathcal{R}} &\simeq \mathcal{D}\left\{\frac{f^{(3)}_{\lambda\mathcal{R}\mathcal{R}}\left( s_1 k_{1}, s_1 k_{1}, s_1 k_{1}\right)}{f_{2,\lambda}\left(\tilde{s}_1 k_{1}\right)^{3/2}}+\left[\frac{f^{(3)}_{\lambda\mathcal{R}\mathcal{R}}\left(s_2 k_{2}, s_2 k_{2}, s_2 k_{2}\right)}{f_{2, \mathcal{R}}\left(\bar{s}_2 k_{2}\right)^{3/2}}+ k_2 \to k_3\right]\right\} \\\nn 
&\quad\quad\quad\quad\quad\quad\quad\quad\quad\quad\quad\quad\quad\quad\quad\quad\quad\quad\quad\quad\quad\quad\quad \times \prod_{i=2,3} \left[ f_{2,\lambda}\left(\tilde{s}_1 k_{1}\right) f_{2,\mathcal{R}}\left(\bar{s}_2 k_{i}\right)\right]^{1/2},
\end{align}
where $s_{1,2} = x^c_{3,\mathcal{\lambda\mathcal{R}\mathcal{R}}}/k^{\rm m}_{1,2}$, $\tilde{s}_{1} = x^c_{2,\lambda}/k^{\rm m}_{1}$ and $\bar{s}_{2} = x^{c}_{2,\mathcal{R}}/k^{\rm m}_{2}$.  Using the Tables \ref{tab:fit1} and \ref{tab:f3fit} for a given set of model parameters $\xi_*$ and $\delta$, one can fix the overall coefficient $\mathcal{D}$ by matching the approximate expression \eqref{f3mrrapp} at its maximum to the exact bispectrum at the triangle configuration where it peaks, \ie $f^{3,\rm max}_{\lambda\mathcal{R}\mathcal{R}} = f^{(3)}_{\lambda\mathcal{R}\mathcal{R}}(k^{\rm m}_1,k^{\rm m}_2)$.
For $\delta = 0.3$ and $\xi_* =5$, the accuracy of \eqref{f3rmmapp} and \eqref{f3mrrapp} derived through the procedure we described above is shown in the top and bottom panels of Figure \ref{fig:S}. Since this process does not require a specific choice of the model parameters, we anticipate that it will also generate accurate factorized forms of the mixed bispectra for other choices of model parameters $\delta,\xi_*$. 

\end{appendix}

\addcontentsline{toc}{section}{References}
\bibliographystyle{utphys}
\bibliography{paper2}
\end{document}